\documentclass[paperpaper,superscriptaddress,english,floatfix, showpacs, amsfonts, amssymb]{revtex4}
\usepackage{epsfig,graphicx,psfrag,amsmath,amssymb,float}
\usepackage[T1]{fontenc}
\usepackage[latin9]{inputenc}
\usepackage{amsmath}
\usepackage{amssymb}
\usepackage{bbold}
\usepackage{amscd}
\usepackage{bm}
\usepackage{psfrag}
\usepackage{epsfig}
\usepackage{graphicx}
\usepackage{dcolumn}
\usepackage{bm}
\usepackage{subfigure}

\usepackage{babel}

\begin{document}
\title{Absence of high-temperature ballistic transport in the spin-$1/2$ $XXX$ chain within the
grand-canonical ensemble}

\author{J. M. P. Carmelo}
\affiliation{Department of Physics, University of Minho, Campus Gualtar, P-4710-057 Braga, Portugal}
\affiliation{Center of Physics of University of Minho and University of Porto, P-4169-007 Oporto, Portugal}
\affiliation{Beijing Computational Science Research Center, Beijing 100193, China}
\author{T. Prosen}
\affiliation{Department of Physics, FMF, University of Ljubljana, Jadranska 19, 1000 Ljubljana, Slovenia}

\date{30 August 2016}

\begin{abstract}
Whether in the thermodynamic limit, vanishing magnetic field $h\rightarrow 0$, and nonzero temperature the spin stiffness of the spin-$1/2$ $XXX$ 
Heisenberg chain is finite or vanishes within the grand-canonical ensemble remains an unsolved and controversial issue, as different 
approaches yield contradictory results. Here we provide an upper bound on the stiffness and show that within that ensemble it 
vanishes for $h\rightarrow 0$ in the thermodynamic limit of chain length $L\to\infty$, at high temperatures $T\rightarrow\infty$.
Our approach uses a representation in terms of the $L$ physical spins $1/2$. For all configurations that generate the
exact spin-$S$ energy and momentum eigenstates such a configuration involves a number $2S$ of unpaired spins $1/2$ in multiplet configurations and
$L-2S$ spins $1/2$ that are paired within $M_{\rm sp}=L/2-S$ spin-singlet pairs. The Bethe-ansatz strings of length $n=1$
and $n>1$ describe a single unbound spin-singlet pair and a configuration within which $n$ pairs are bound, respectively.
In the case of $n>1$ pairs this holds both for ideal and deformed strings associated with $n$ complex rapidities with the 
same real part. The use of such a spin $1/2$ representation provides useful physical information on the problem under 
investigation in contrast to often less controllable numerical studies. Our results provide strong evidence for the absence 
of ballistic transport in the spin-$1/2$ $XXX$ Heisenberg chain in the thermodynamic limit, for high temperatures $T\to\infty$, vanishing 
magnetic field $h\rightarrow 0$ and within the grand-canonical ensemble.
\end{abstract}

\pacs{75.10.Pq, 75.40.Gb, 72.25.-b, 75.76.+j}

\maketitle

\section{Introduction}
\label{Introduction}

The anisotropic spin-$1/2$ $XXZ$ Heisenberg chain \cite{Taka-AN} with anisotropy parameter $\Delta\ge 0$, exchange 
integral $J$, and Hamiltonian,
$J\sum_{j=1}^{L}({\hat{S}}_j^{x}{\hat{S}}_{j+1}^{x} + {\hat{S}}_j^{y}{\hat{S}}_{j+1}^{y} + \Delta\,{\hat{S}}_j^z{\hat{S}}_{j+1}^z)$,
where $\hat{S}_j^{x,y,z}$ are components of the spin-$1/2$ operators at site $j=1,...,L$, is a paradigmatic example of an 
integrable strongly correlated quantum many-body system. 

However, the isotropic point at $\Delta=1$ (the spin-$1/2$ $XXX$ Heisenberg chain \cite{Bethe,Taka})
is the most experimentally relevant \cite{SPA-11,Motoyama-96,Thurber-01}.
It is also the case that poses the most challenging technical problems for theory. 
For instance, the problem of clarifying the possibility of ballistic spin transport at nonzero temperatures in the 
spin-$1/2$ $XXX$ chain in a magnetic field $h$ is one of the most intensely debated unsettled fundamental questions 
in the theory of strongly correlated systems. Its Hamiltonian with periodic boundary conditions reads,
\begin{equation}
\hat{H} = J\sum_{j=1}^{L} \hat{\vec{S}}_j \cdot \hat{\vec{S}}_{j+1}- 2\mu_B\,h\sum_{j=1}^{L} \hat{S}_j^z \, ,
\label{H-h}
\end{equation}
where $h \in [-h_c,h_c]$, $\mu_B$ is the Bohr magneton and $\pm h_c=\pm J/\mu_B$ are the critical fields for fully polarized 
ferromagnetism. 

The model's spin stiffness $D(T)$, also called spin
Drude weight, defined via the singularity in the real part of the spin conductivity, 
\begin{equation}
\sigma (\omega,T) = 2\pi\,D (T)\,\delta (\omega) +  \sigma^{reg} (\omega,T) \, ,
\end{equation}
can be interpreted as a quantitative measure of ballistic spin transport.
In the thermodynamic limit (TL), $L\rightarrow\infty$, the corresponding stiffness expressions given below in this paper 
involve the expectation values of the $z$-component spin current operator, 
\begin{equation}
\hat{J}^z = -i\,J\sum_{j=1}^{L}(\hat{S}_j^+\hat{S}_{j+1}^- - \hat{S}_{j+1}^+\hat{S}_j^-)  \, ,
\label{Jz-current}
\end{equation}
where $\hat{S}_j^{\pm} = \hat{S}_j^x \pm i \hat{S}_j^y$.

Different approximate approaches \cite{SPA-11,ZNP-97,Peres-99,Zotos-99,Peres-00,Gros-02,Kawa-03,Cabra-03,PSZL-04,ANI-05,Shastry-08,SPA-09,Tomaz-11,Znidaric-11,HPZ-11,Karrasch-12,Karrasch-13,Znidaric-13,Tomaz-13,Robin-14,CPC}, ranging from numerical simulations 
through effective field-theoretical descriptions to calculations partially based on the Bethe ansatz (BA) have 
yielded different, contradictory results, either showing that the model's spin stiffness $D(T)$
converges as $h\rightarrow 0$ in the TL to zero \cite{SPA-11,Zotos-99,Peres-99,HPZ-11} or to a finite value 
\cite{Karrasch-13,Karrasch-12,ANI-05,Cabra-03,Gros-02}.  

For instance, the schemes used in the studies of
Refs. \cite{Karrasch-13,Karrasch-12,ANI-05,Cabra-03,Gros-02} lead to a finite value for the spin stiffness
at nonzero temperature. In contrast, the investigations of Ref. \cite{SPA-11} indicate that transport
at finite temperatures is dominated by a diffusive contribution, the spin stiffness being very small or zero.
Such studies exclude the large spin stiffness found in Ref. \cite{ANI-05} by a phenomenological
method that relies on a spinon and anti-spinon basis for the thermodynamic Bethe ansatz (TBA) \cite{Taka}. 
The results obtained by a completely different and more direct use of the TBA in Refs. \cite{Zotos-99,Peres-99} 
as well as the more recent results of Ref. \cite{HPZ-11} that rely on the combination of several techniques
find a vanishing spin stiffness for zero spin density.

The nature of the exotic spin transport properties at nonzero temperature 
of one-dimensional (1D) correlated lattice systems has been a problem of also
experimental interest \cite{Steiner-76,Motoyama-96,Thurber-01,Pratt-06,Branzoli-11,Yang-12,Yager-13}.
The spin stiffness is directly related to the long-time asymptotic current-current correlation function as,
\begin{equation}
D (T) = \frac{1}{2LT} \lim_{t\rightarrow\infty}\langle \hat{J}^z (t) \hat{J}^z (0)\rangle \, .
\end{equation} 
(The angle brackets $\langle . \rangle$ denote here the thermal average.) In integrable models there is a lower bound for $D (T)$, which is encoded 
in an inequality due to Mazur \cite{Mazur}, 
\begin{equation}
D (T) \geq \frac{1}{2L}\sum_{j}{\langle \hat{J}^z\hat{Q}_{j}\rangle^2\over \langle \hat{Q}_j^2\rangle} \, .
\end{equation}
Here the sum runs over a complete set of {\em linearly extensive} orthogonal commuting conserved quantities $\hat{Q}_j$ 
for which $\langle{\hat{Q}_j^2}\rangle\propto L$, local and quasilocal \cite{Tomaz-11,Tomaz-13,Suzuki,Tomaz-14,Pereira}. In the case of
the spin-$1/2$ $XXZ$ chain, the sum over  strictly {\it local} conserved quantities responsible for integrability gives at
nonzero temperatures (i) a finite value and thus ballistic spin transport for $h\neq 0$ and (ii) vanishes
and is inconclusive at $h=0$.

Two recent results provided some essential preliminary steps for the clarification of the problem studied in this paper.
The first of these results is that the Mazur's inequality sum over {\it quasilocal} 
conservation laws associated with deformed symmetries gives for the spin-$1/2$ $XXZ$ chain a stiffness lower bound at $h=0$,
$D_l (T) \leq D (T)$, which for $T\rightarrow\infty$ reads \cite{Tomaz-13},
\begin{equation}
D_l (T) = {16 J^2 \over T}{\sin^2 (\pi l/l')\over\sin^2 (\pi/l')}
\left(1 - {l'\over 2\pi}\sin \left({2\pi\over l'}\right)\right) \, . 
\label{Dl}
\end{equation}
It refers to a dense set of commensurate easy-plane anisotropies, $\Delta = \cos (\pi l/l')$,
where $l$, $l'\in\mathbb{Z}^+$ and $l\leq l'>0$ are such that $0\leq \Delta\leq 1$. 
Since this lower bound vanishes at the isotropic point, $\Delta = 1$, it does
not discard the possibility that the spin stiffness of the spin-$1/2$ $XXX$ chain is also vanishing as $h\rightarrow 0$.

\begin{figure}
\begin{center}
\centerline{\includegraphics[width=7.00cm]{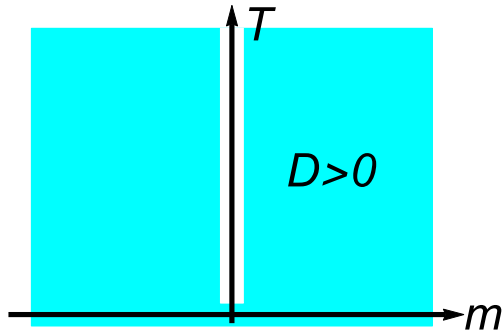}}
\caption{Phase diagram of ballistic spin transport of the spin-$1/2$ $XXX$ Heisenberg chain. Ballistic regions with positive 
spin Drude weight, $D>0$, namely temperature $T=0$ or spin density $m \ne 0$, are painted in cyan, whereas in the 
complementary region, $T>0$ and $m=0$ (white), the spin stiffness vanishes, $D=0$, in the thermodynamic limit.}
\label{figure1}
\end{center}
\end{figure}

The second recent result presented in Ref. \cite{CPC} is a upper bound for the spin stiffness of the 
spin-$1/2$ $XXX$ chain, $D_u (T) \geq D (T)$, valid within the canonical ensemble
for spin densities $m\in [0,1]$ and the whole $T>0$ range, in the TL. Its limiting behaviors are, 
\begin{eqnarray}
D_{u} (T) & = & \frac{(J\pi)^2}{2T}\,m^2\,L \, , \hspace{0.25cm}{\rm for}\hspace{0.20cm}m \ll 1 \, ,
\nonumber \\
& = & \frac{J^2}{2T}\, (1-m)^2\,L  \, , \hspace{0.25cm}{\rm for}\hspace{0.20cm}(1-m)\ll 1 \, .
\label{value-allT-D-m0-1}
\end{eqnarray}
That $D_{u} (T)$ vanishes as $m^2\,L$ in the $m\rightarrow 0$ limit ensures that within the canonical ensemble the 
stiffness vanishes as $m\rightarrow 0$ yet leaves out, marginally, the grand canonical ensemble as $h\rightarrow 0$ in which 
$\langle m^2\rangle = {\cal O}(1/L)$. A schematic phase diagram of temperature $T$ versus spin
density $m$ of ballistic spin transport is shown in Fig. \ref{figure1}.

In this paper we provide new insights on the above unsolved problem concerning the spin stiffness for
the spin $1/2$ $XXX$ chain in the TL. Specifically, we provide strong theoretical evidence that for 
high temperatures $T\rightarrow\infty$ it also vanishes for $h\rightarrow 0$, within the grand-canonical ensemble.
While for a canonical ensemble one considers that the spin density $m$ is kept constant, in the case of a grand-canonical ensemble
it is the magnetic field $h$ that is fixed. In general the canonical-ensemble and grand-canonical ensemble lead to
the same results in the TL. This is generally true {\it except} near a phase transition or a critical point. Hence
this issue deserves a careful analysis in the $m\rightarrow 0$ and $h\rightarrow 0$ limits, respectively.

The use of effective spinon representations \cite{Faddeev,Talstra-97,Enderle-10} provides a suitable description of the model 
low-energy physics and excitations of the $S=0$ ground state. However, they do not apply to high-temperature problems
at a magnetic field $h \in [-h_c,h_c]$ that involve all $2^L$ energy eigenstates, as that studied in this paper. Our approach 
then rather uses the representation of Ref. \cite{CPC} in terms of the spin-$1/2$ $XXX$ chain $L$ physical spins $1/2$. 
Within such a representation, all configurations that generate the 
exact energy and momentum eigenstates of spin $S$ involve a number $2S$ of unpaired spins $1/2$ 
in multiplet configurations and $L-2S$ spins $1/2$ that are paired within 
$M_{\rm sp}=L/2-S$ spin-singlet pairs. Within the TBA, the imaginary part of the complex rapidities simplify
in the TL, which corresponds to the ideal strings of length $n>1$ \cite{Taka}. For large $L$ values there is in addition
two types of deformed complex rapidities that deviate from such an ideal behavior \cite{Takahashi-03,Essler-92,Vladimirov-84}.

Importantly, the general representation in terms of $2S$ unpaired physical spins $1/2$ plus $M_{\rm sp}=L/2-S$ spin-singlet pairs
of physical spins $1/2$ used in the studies of this paper applies both to the TBA \cite{Taka} and to BA schemes including
three types of complex rapidities \cite{Takahashi-03}, respectively. On the one hand, both for an ideal string and a deformed 
string of length $n>1$  the corresponding set of $n$ complex rapidities with the same real part refer to an independent 
configuration with a number $n$ of spin-singlet pairs bound within it. On the other hand,
the real rapidities correspond to single unbound spin-singlet pairs. 

Our derivation relies on the spin stiffness expression in terms of matrix elements of the $z$-component current operator,
Eq. (\ref{Jz-current}), and the operator algebra relating that operator to both the other two $SU(2)$ 
symmetry operator components,
\begin{equation}
\hat{J}^{+} = (\hat{J}^{-})^{\dag} = 2i\,J\sum_{j=1}^{L}(\hat{S}_j^+\hat{S}_{j+1}^z - \hat{S}_{j+1}^+\hat{S}_j^z) \, ,
\label{Jpm}
\end{equation}
and the three generators ${\hat{S}}^{\eta} = \sum_{j=1}^{L}{\hat{S}}_j^{\eta}$, $\eta=\pm,z$ of that global symmetry. This includes
the commutators, 
\begin{eqnarray}
\left[\hat{J}^z,\hat{S}^{\pm}\right] & = &
\left[\hat{S}^{z},\hat{J}^{\pm}\right] = \pm \hat{J}^{\pm} 
\, ; \hspace{0.25cm}
\left[\hat{J}^{\pm},\hat{S}^{\mp}\right] = \pm 2\hat{J}^z
\nonumber \\
\left[\hat{J}^z,\hat{S}^{z}\right] & = & 0
\, ; \hspace{0.25cm}
\left[\hat{J}^z,(\hat{\vec{S}})^2\right]  = \hat{J}^{+}\hat{S}^{-} - \hat{S}^{+}\hat{J}^{-} \, ,
\label{comm-currents}
\end{eqnarray}
which follow directly from the $SU(2)$ algebra for the operators under consideration. 

There is a general consensus that the use of ideal strings of TBA for energy and momentum eigenstates described by groups
of real and complex rapidities \cite{Taka} leads in the TL to exact results as long as either the temperature 
or the magnetic field are nonzero \cite{Takahashi-03,Wiegmann-83}. 
Our studies involve the spin stiffness at very hight temperature, $T\rightarrow \infty$, so that concerning thermal effects
they are not affected in the TL by the string deformations. On the one hand, concerning the case $h=0$, we use a method 
other than the BA or TBA to compute the exact current operator expectation values
of the corresponding $S^z=0$ energy and momentum eigenstates \cite{CPC}.
On the other hand, in what the contributions to the spin stiffness for the
model at finite magnetic field from the square of current operator expectation 
values of finite-$S^z$ energy and momentum eigenstates is concerned, we rely on upper bounds. In contrast
to those used in Ref. \cite{CPC}, the present upper bounds involve sums that run over a large, macroscopic number, of energy 
and momentum eigenstates. As justified below in Sec. \ref{stringHTh0}, such upper bounds are in the TL insensitive 
to the use of ideal \cite{Taka} or deformed \cite{Takahashi-03} BA strings. 

Our representation in terms of configurations of the $L$ physical spins $1/2$ provides useful physical 
information on the problem under investigation, in contrast to the often less controllable numerical studies on the 
occurrence or lack of ballistic spin transport in the spin-$1/2$ $XXX$ chain as $h\rightarrow 0$ in the TL.

The remainder of the paper is organized as follows. In Sec. \ref{stiffnessAT} the finite-temperature spin 
stiffness and the representation in terms of configuration of the $L$ physical spins $1/2$ used in the studies of 
this paper are introduced. The general expressions of the spin stiffness at high temperature $T\rightarrow\infty$ is the 
issue addressed in Sec. \ref{stiffnessHT}. In Sec. \ref{stiffnessHTh0only} a non-BA-related method used 
to compute the spin currents of the $S^z=0$ energy and momentum eigenstates for the strictly zero magnetic-field 
case is briefly reported and the physical consequences of the corresponding exact results are discussed. 
Useful and needed inequalities and corresponding current absolute values upper bounds are introduced in Sec. \ref{CurrentsUB}. 
The effects of the string deformations on the spin currents in the TL at finite magnetic field is the issue addressed in 
Sec. \ref{stringHTh0}. In Sec. \ref{stiffnessHTh0} the high-temperature stiffness upper bounds within the TL 
used in our study are derived. Finally, the concluding remarks are presented in Sec. \ref{concluding}. Additional
technical information useful for details of our analysis is provided in Appendices \ref{ncurrents} and \ref{currentUB2}.

\section{The finite-temperature spin stiffness and $L$ physical spins $1/2$ configurations}
\label{stiffnessAT}

We denote the energy eigenstate's spin and spin projection by $S$ and
$S^{z}=-(N_{\uparrow} -N_{\downarrow})/2$, respectively. Here $N_{\uparrow}$ and $N_{\downarrow}$
such that $L = N_{\uparrow} + N_{\downarrow}$ are the numbers of spins $1/2$ with up and down spin
projection, respectively. For the so-called lowest-weight-states (LWSs) and highest-weight-states (HWSs) of the 
$SU(2)$ algebra we have $S=-S^z$ and $S=S^z$, respectively. The class of LWSs and the non-LWSs 
generated from those that are used in our analysis are energy and momentum eigenstates.
They are as well eigenstates of $(\hat{\vec{S}})^2$ and $\hat{S}^z$ with eigenvalues $S(S+1)$ and $S^z$,
respectively. We thus label all 
$2^{L}$ energy, momentum (as well as spin and spin projection) eigenstates by $\vert l_{\rm r},S,S^z\rangle$. Here $l_{\rm r}$ stands for all quantum 
numbers other than $S$ and $S^z$ needed to specify an energy and momentum eigenstate, $\vert l_{\rm r},S,S^z\rangle$. 
This is independent of using the general BA or the TBA for these states, always holding that
$\sum_{l_{\rm r}}={\cal{N}}_{\rm singlet} (S)$ for the model in each fixed-$S$ subspace. Here
${\cal{N}}_{\rm singlet} (S) = {L\choose L/2-S}-{L\choose L/2-S-1}$ is that subspace number of independent 
spin-singlet configurations and thus ${\cal{N}} (S)=(2S+1)\,{\cal{N}}_{\rm singlet} (S)$ is its dimension.
Since the LWSs and non-LWSs generated from them considered in this paper are energy and momentum eigenstates, 
these designations are often used for the latter states.

Within the canonical-ensemble description at fixed value of $S^z$, the spin stiffness $D (T)$ expression
involves the current operator expectation values, $\langle l_{\rm r},S,S^z\vert\hat{J} \vert l_{\rm r},S,S^z\rangle$, which 
{\it in the TL and for nonzero temperatures} are the current matrix elements that contribute to it \cite{ZNP-97,CPC,Ginsparg-90}. 
As justified below in Sec. \ref{stiffnessHTh0only}, for the non-LWSs, which are generated from the corresponding
LWSs $\vert l_{\rm r},S,-S\rangle$ as 
$\vert l_{\rm r},S,S^z\rangle = 
\frac{1}{\sqrt{{\cal{C}}}}({\hat{S}}^{+})^{n_s}\vert l_{\rm r},S,-S\rangle$
where ${\cal{C}} = (n_s!)\prod_{j=1}^{n_s}(\,2S+1-j\,)$ and $n_s\equiv S + S^z = 1,...,2S$,
such current operator expectation values can be expressed in terms of that of the corresponding LWS
by suitable use of the spin $SU(2)$ operator algebra. From such considerations one finds that 
in the TL the spin stiffness reads $D (T) = 0$ for $S^z = 0$ and for $\vert S^z\vert \geq  1/2$ it 
can be written as \cite{CPC},
\begin{equation}
D (T) = {(2S^z)^2\over 2 L T}\sum_{S=\vert S^z\vert}^{L/2}\sum_{l_{\rm r}} 
p_{l_{\rm r},S,S^z}{\vert\langle\hat{J}^z (l_{\rm r},S)\rangle\vert^2\over (2S)^2} \, .
\label{D-all-T-simp}
\end{equation}
Here $\hat{J}^z$ is the $z$ component of the spin current operator, Eq. (\ref{Jz-current}), 
$p_{l_{\rm r},S,S^z}$ are the Boltzmann weights, and $\langle\hat{J}^z (l_{\rm r},S)\rangle\equiv\langle l_{\rm r},S,-S\vert\hat{J} \vert l_{\rm r},S,-S\rangle$
are the LWSs spin currents. In this and all following expressions for the spin stiffness, 
the sums over $S$ always increase in  {\em steps} of $1$, whereas $S^z$ and
$S$ have to be integers (half-odd integers) for even (odd) $L$. 

For each $S$ value there are ${\cal{N}}(S) = (2S+1)\,{\cal{N}}_{\rm singlet} (S)$ energy and momentum 
eigenstates. Our study accounts for all corresponding $\sum_{2S=0\,({\rm integers})}^{L}\,{\cal{N}}(S) = 2^{L}$ 
energy and momentum eigenstates. For $S>0$ each such a state is populated by a set of $2S$ spins $1/2$ that
participate in its multiplet configuration, which is one of the $2S+1$ multiplet configurations, 
and a complementary set of even number $L-2S$ of
spins $1/2$ that form a tensor product of singlet states. Since all the ${\cal N}(S)$ states with the same $S$ value
have the same $\hat{\vec{S}}^2$ eigenvalue, the energy and momentum eigenstates are superpositions of such configuration terms.
Each such terms is characterized by a different partition of $L$ physical spins $1/2$ into $2S$ such spins 
that participate in a $2S+1$ spin multiplet and a product of singlets involving the remaining
even number $L-2S$ of spins $1/2$. 

As in Ref. \cite{CPC}, we call {\it unpaired spins} and {\it paired spins} the 
members of such sets of $2S$ and $L-2S$ spins, respectively. In the TL this partition is common to
the general BA solution and the TBA representation of its energy and momentum eigenstates. 
Both for large $L$ and within the TL the $L-2S$ paired spins $1/2$ are contained in a number,
\begin{equation}
M_{\rm sp} = {1\over 2}(L-2S) = {L\over 2}(1-m_S) \, ,
\label{Nsingletpairs0}
\end{equation}
of spin-singlet pairs. Hence each fixed-$S$ subspace is spanned by energy and momentum
eigenstates with {\it exactly} the same number $M_{\rm sp} = L/2-S$ of such pairs.
Moreover, $M_{\rm sp} = L/2-S$ also is the total number of BA rapidities that
describe such states. And this is independent of such rapidities being all real or
some being real and and other complex. Consistently, within the present
representation each BA rapidity describes a spin-singlet pair.

The derivation of the spin stiffness upper bound of Ref. \cite{CPC}, whose limiting behaviors are given
in Eq. (\ref{value-allT-D-m0-1}), used a large overestimate of the current absolute values 
$\vert\langle\hat{J}^z (l_{\rm r},S)\rangle\vert$. Specifically, for the whole set of energy and momentum 
eigenstates with the same $S^z$ value corresponding to the sums $\sum_{l_{\rm r} } \sum_{S=\vert S^z\vert}^{L/2}$ 
in Eq. (\ref{D-all-T-simp}) it used the largest magnitude of the current expectation value among 
these states. Since the probability distribution $p_{l_{\rm r},S,S^z}$ in each fixed-$S^z$ canonical 
ensemble is normalized as $\sum_{S=\vert S^z\vert}^{L/2}\sum_{l_{\rm r}} p_{l_{\rm r},S,S^z}=1$,
this then allowed performing exactly such sums for all nonzero temperatures, $T>0$. 

The large overestimate of the currents used in deriving that spin stiffness upper bound 
is behind its $m\rightarrow 0$ behavior reported in Eq. (\ref{value-allT-D-m0-1}) leaving out the grand canonical 
ensemble in which $\langle m^2\rangle = {\cal O}(1/L)$. Our main goal is to derive an alternative spin
stiffness upper bound whose estimate of the current absolute values $\vert\langle\hat{J}^z (l_{\rm r},S)\rangle\vert$ is closer to 
yet larger than those of the currents in Eq. (\ref{D-all-T-simp}). Here we perform such
a program for high temperatures, $T\rightarrow\infty$. 

The $M_{\rm sp} = L/2-S$ spin-singlet pairs of
each energy and momentum eigenstate include $M^p$ 
unbound pairs. The remaining $M_{\rm sp}^B=M_{\rm sp}-M^p$ spin-singlet pairs
of energy and momentum eigenstates described by groups of both real and complex BA rapitities
are bound within a well-defined number $M_{\rm st}^B<M_{\rm sp}^B$ of independent configurations.
(For energy and momentum eigenstates described only by groups of real BA rapitities
such numbers read $M^p=M_{\rm sp}$ and $M_{\rm sp}^B=0$, respectively.)
As discussed in the following, there is a one-to-one correspondence between 
such $M_{\rm st}^B$ independent configurations and the $M_{\rm st}^B$ strings
of length larger than one, each of which is associated with a set of
complex BA rapidities with the same real part. 

The unbound and bound spin-singlet pairs of the $L-2S$ paired spins
are indeed described by groups of real and complex solutions, respectively,  of the model 
BA equation \cite{Bethe,Taka},
\begin{equation}
2\arctan (\Lambda_j) = q_j + {1\over L}\sum_{\alpha\neq j}2\arctan\left({\Lambda_j\!-\!\Lambda_{\alpha}\over 2}\right)\!\!\!\!\pmod{2\pi}.
\label{gen-Lambda-BA}
\end{equation}
Here the $\alpha =1,...,M^p$ summation 
is over the subset of occupied $q_{\alpha}$ quantum numbers out of the full set,
\begin{equation}
q_j = {2\pi\over L}I_j \, , \hspace{0.20cm}{\rm where}\hspace{0.20cm} j=1,...,M^b \, ,
\label{qjR}
\end{equation}
$M^b = M^p+M^h$, and $M^h=2S + 2(M_{\rm sp}^B - M_{\rm st}^B)$. The different occupancy configurations of the related quantum 
numbers $I_j$ (defined modulo $L$) such that $j = 1,...,M^b$ generate different energy and momentum 
eigenstates. The latter are successive integers or half-odd integers according to the boundary conditions,
\begin{eqnarray}
I_j & = & 0,\pm 1,...,\pm {M^b -1\over 2} \hspace{0.20cm}{\rm for}\hspace{0.20cm}M^b\hspace{0.2cm}{\rm odd} \, ,
\nonumber \\
& = & \pm 1/2,\pm 3/2,...,\pm {M^b -1\over 2} \hspace{0.20cm}{\rm for}\hspace{0.20cm}M^b\hspace{0.2cm}{\rm even} \, .
\label{Ij}
\end{eqnarray}

The set of $j=1,...,M^b$ quantum numbers $q_j$ can only have occupancy zero and one, respectively.
Within our representation, the $\alpha =1,...,M^p$ occupied momentum values 
$q_{\alpha}$ refer to the center of mass translation degrees of freedom of $M^p$ neutral composite 
{\it pseudoparticles}. The internal degrees of freedom of each of these
$M^p$ pseudoparticles refer to one of the $M^p$ unbound spin-singlet pairs. 

Our functional representation involves a $q_j$ distribution function $M^p (q_{j})$ that reads $0$ and $1$ for the 
$M^h=2S + 2(M_{\rm sp}^B - M_{\rm st}^B)$ unoccupied and $M^p$ occupied $q_j$ values, respectively. Since the
contribution to the momentum eigenvalues of the $M^p$ pseudoparticles
reads $\pi + \sum_{j=1}^{M^b}M^p (q_{j})\,q_j$, the set $j = 1,...,M^b$ 
of quantum numbers $q_j$ such that $q_{j+1}-q_j=2\pi/L$ may be associated with the discrete 
momentum values of a pseudoparticle spin band. For LWSs described only by groups of real rapidities, 
all $M_{\rm sp} = L/2-S$ spin-singlet pairs are unbound, so that
$M^p = M_{\rm sp} = L/2-S$, $M^h=2S$, $M^b = L/2+S$, $M_{\rm sp}^B=0$, and $M_{\rm ps}^B=0$.

Consistently with the $0$ and $1$ allowed occupancies of the
spin-band momentum values, the LWS BA wave functions formally vanish when two rapidities $\Lambda_j$ and $\Lambda_{j'}$ 
in Eq. (\ref{gen-Lambda-BA}) become equal. If one considered all the rapidities to be real, this property could suggest that simply 
choosing $\alpha =1,...,M^p$ distinct occupied momentum values
$q_{\alpha}$ among the set of $j = 1,...,M^b$ allowed spin-band discrete momentum values $q_j$, which gives a dimension
${M^b\choose M^p}={L/2+S\choose L/2-S}$, would allow the reconstruction of all $2^L$ energy eigenstates that span the model Hilbert space. 

However, only some of the solutions to the model general BA equation involve only a group of 
$M_{\rm sp}=M^p$ real rapidities $\Lambda_j$. As mentioned above, there also exist solutions 
involving groups of real and complex rapidities \cite{Bethe,Taka}. 
There are $M_{\rm sp}=M^p+M_{\rm sp}^B$ BA rapidities that describe the $M_{\rm sp}$ spin-singlet pairs of a 
general energy and momentum eigenstate. Within our representation in terms of $L-2S$ 
paired physical spins $1/2$, the $M^p$ real rapidities and $M_{\rm sp}^B$ complex rapidities 
describe their $M^p$ unbound spin-singlet pairs and their $M_{\rm sp}^B$ spin-singlet pairs
bound within the state $M_{\rm st}^B$ independent configurations, respectively. 

The following general relations between the different numbers under consideration apply,
\begin{eqnarray}
M^p & = & M_{\rm sp} - M_{\rm sp}^B = {L\over 2}(1-m_S) - M_{\rm sp}^B \, ,
\nonumber \\
M^h & = & 2S + 2(M_{\rm sp} - M_{\rm st}) = 2S + 2(M_{\rm sp}^B - M_{\rm st}^B) \, ,
\nonumber \\
M^b & = & M^p + M^h \, .
\label{MpMhMbcomplex}
\end{eqnarray}
Here $M_{\rm st} = M^p + M_{\rm st}^B$ gives the total number of both $M^p$ unbound spin-singlet
pairs and corresponding spin-band pseudoparticles and $M_{\rm st}^B$ independent
$n$-pair configurations with $n>1$ spin-singlet pairs bound within them. The $n$ complex rapidities
with the same real part that describe each such a $n$-pair configuration is labelled 
by a quantum number $l = 1,...,n$. It also labels each of
the spin-singlet pairs bound within such a configuration. These $l = 1,...,n$ rapidities 
with the same real part have the general form \cite{Takahashi-03},
\begin{equation}
\Lambda_j^{n,l} = \Lambda_j^n + i (n+1-2l) + D_j^{n,l} \hspace{0.25cm}{\rm where}\hspace{0.20cm} l = 1,...,n \, .
\label{Lambda-jnl}
\end{equation}
The roots of Eq. (\ref{gen-Lambda-BA}) are here partitioned in a configuration of strings. A $n$-string is a group
of $n$ roots also called rapidities. Within our representation such a string describes an independent $n$-pair configuration.
The number $n$ is often called the string length. The real part of the set of $n$ rapidities, $\Lambda_j^n$, 
is called the string center \cite{Takahashi-03}. Hence $M_{\rm st} = M^p + M_{\rm st}^B$ is in Eq. (\ref{MpMhMbcomplex})
the number of strings. $M^p$ and $M_{\rm st}^B$ refer to the number of strings of length $n=1$ and
length $n>1$, respectively. Note that for $n=1$ one has that $l=1$ and the corresponding single
rapidity $\Lambda_j^{1,1}$ is real. The quantity $D_j^{n,l} = R_j^{n,l} + i \delta_j^{n,l}$ in Eq. (\ref{Lambda-jnl}), where $R_j^{n,l}$ and $\delta_j^{n,l}$ 
are real numbers, is the fine-structure deviation from the TBA ideal strings for which $D_j^{n,l}=0$ \cite{Taka}. 
Importantly, $D_j^{1,1}=0$ for the $M^p$ real rapidities $\Lambda_j^{1,1}$ of all energy and momentum eigenstates.

There is a one-to-one correspondence between an energy eigenstate $M_{\rm st}^B$
strings of length $n>1$ and the $M_{\rm st}^B$ independent $n$-pair configurations with
$n>1$ spin-singlet pairs bound within them, respectively.
The string length $n>1$ is thus the number of spin-singlet pairs bound within the corresponding $n$-pair configuration.
The present representation clarifies the physical meaning of the imaginary parts of the $n>1$ complex rapidities 
with the same real part that refer to a string of length $n$, Eq. (\ref{Lambda-jnl}): Such imaginary
parts are associated with the binding within the corresponding $n$-pair configuration of $n>1$ spin-singlet pairs.
Consistently and as mentioned above, for $n=1$ the rapidity $\Lambda_j^{1,1}$ is real and describes a single unbound pair.

The maximum possible value of the number $n$ of spin-singlet pairs bound within a 
$n$-pair configuration and corresponding string length is obviously given by the number
of spin-singlet pairs, $M_{\rm sp}=(L-2S)/2$, Eq. (\ref{Nsingletpairs0}).  The set of
energy and momentum eigenstates that span each fixed-$S$ subspace have all the
same number $M_{\rm sp}=(L-2S)/2$ of such pairs. Provided that $(1-m_s)$ is finite,
that number is such that $M_{\rm sp}\rightarrow\infty$ as $L\rightarrow\infty$.
Hence in general we consider in the TL that $n$ has the range $n = 1,...,\infty$. 

For a given large $L$, the complex solutions of the spin-$1/2$ $XXX$ chain BA equation, Eq. (\ref{gen-Lambda-BA}),
are found to belong to three classes \cite{Takahashi-03}. The first class refers to the ideal strings
for which $D_j^{n,l}=0$ in Eq. (\ref{Lambda-jnl}). The second class was first identified by Essler, Korepin and Schoutens
(EKS) for $n=2$ complex rapidities \cite{Essler-92} yet also occurs for $n>2$. The corresponding strings deviate from
the ideal behavior and are known as EKS-strings \cite{Takahashi-03}. The imaginary part of their complex
rapidities are smaller than $1/2$. It decreases upon increasing $L$, vanishing at some $L$ value. The 
third class of solutions corresponds to another type of deformed strings usually called V-strings, which have been 
first found by Vladimirov (V) \cite{Vladimirov-84}. In the case of a system with a fixed large $L$, the number of
energy eigenstates obtained by accounting for the three classes of BA equations groups of real and complex solutions 
is given by the correct Hilbert space dimension, $2^L$ \cite{Takahashi-03}.

In Sec. \ref{stringHTh0} it is justified why concerning the model at finite
magnetic field our final results are independent from the use in the TL of ideal or deformed 
strings of length $n>1$ for the $\vert S^z\vert\geq 1/2$ energy and momentum states described by 
groups of real and complex rapidities. The unbinding of spin-singlet pairs by processes 
associated with the vanishing of the EKS-strings imaginary parts, usually called collapse of narrow 
pairs, is for a large system and finite magnetic field the aberration from the ideal strings that must
be accounted for. The effects of the V-strings are unimportant in the 
TL for the physical quantities studied in this paper. For large finite systems they behave in a rather normal way,
consistent with the predictions of the $1/L$ expansion methods \cite{Takahashi-03}.

The direct relation reported in the following
of the TBA quantum numbers to our representation configurations
of $2S$ unpaired spins $1/2$, $L-2S$ paired spins $1/2$, corresponding $M_{\rm sp} = L/2-S$ 
spin-singlet pairs, and $M^p$ and $M_{\rm sp}^B$ unbound and bound such pairs, respectively,
is useful and needed for the studies of Secs. \ref{CurrentsUB} and \ref{stiffnessHTh0}. 
Within the TBA, the $l = 1,...,n$ complex rapidities of a string, Eq. (\ref{Lambda-jnl}), simplify in the TL to their
ideal form \cite{Taka},
\begin{equation}
\Lambda_j^{n,l} = \Lambda_j^n + i (n+1-2l) \hspace{0.25cm}{\rm where}\hspace{0.20cm} l = 1,...,n \, .
\label{Lambda-jnl-ideal}
\end{equation}
Such rapidities are solutions of the TBA coupled integral equations given below. 
The number $2^L$ of energy eigenstates prevails under the use of the TBA in terms of 
only ideal strings, Eq. (\ref{Lambda-jnl-ideal}).

We call $M_n$ the number of $n$-pair configurations and corresponding strings
of length $n$. Within our representation the $M_{\rm st} = M^p + M_{\rm st}^B$
BA strings correspond to $M_{\rm st} = M^p + M_{\rm st}^B$ $n$-pair configurations involving for each
spin-$S$ energy and momentum eigenstate its $M_{\rm sp}=L/2-S$ spin-singlet pairs, Eq. (\ref{Nsingletpairs0}).
Consistently, the TBA quantum numbers obey the following sum rule \cite{Taka}, 
\begin{equation}
m_{\rm sp} = \sum_{n=1}^{\infty}n\,m_n = {1\over 2}(1-m_S) 
\, ; \hspace{0.5cm}
M_{\rm sp} =  \sum_{n=1}^{\infty}n\,M_n = L/2 - S = m_{\rm sp}\,L \, ,
\label{Nsingletpairs}
\end{equation}
where $m_{\rm sp}$ is the density of spin-singlet pairs and,
\begin{equation}
m_S = 2S/L\geq m \, , \hspace{0.50cm} m_n = M_n/L \, .
\label{mm}
\end{equation} 
 
Within the momentum-distribution functional notation used here and in Ref. \cite{CPC}, the TBA equations 
derived in Ref. \cite{Taka} from the general BA equation, Eq. (\ref{gen-Lambda-BA}), by means of real and
complex rapidities associated with ideal strings, Eq. (\ref{Lambda-jnl-ideal}), read,
\begin{equation}
q_j = k^n_j - {1\over L}\sum_{(n',j')\neq (n,j)}M_{n'} (q_{j'})\,\Theta_{n\,n'} (\Lambda_j^n-\Lambda_{j'}^{n'}) \, .
\label{gen-Lambda}
\end{equation}
In this equation,
\begin{equation}
k^n_j \equiv k^n (q_j) = 2\arctan (\Lambda_j^n/n) \, ,
\label{kn-gen-Lambda}
\end{equation}
and $\Theta_{n\,n'}(x)$ is an odd function of $x$ given by,
\begin{eqnarray}
& & \Theta_{n\,n'}(x) = \delta_{n,n'}\Bigl\{2\arctan\Bigl({x\over 2n}\Bigl) 
+ \sum_{l=1}^{n -1}4\arctan\Bigl({x\over 2l}\Bigl)\Bigr\} 
\nonumber \\
& + & (1-\delta_{n,n'})\Bigl\{ 2\arctan\Bigl({x\over \vert\,n-n'\vert}\Bigl)
+ 2\arctan\Bigl({x\over n+n'}\Bigl) 
\nonumber \\
& + & \sum_{l=1}^{{n+n'-\vert\,n-n'\vert\over 2} -1}4\arctan\Bigl({x\over \vert\, n-n'\vert +2l}\Bigl)\Bigr\} \, .
\label{Theta}
\end{eqnarray}
Here $n, n' = 1,...,\infty$ and $\delta_{n,n'}$ is the usual Kronecker symbol.
(The relation of the $n=1$ rapidity momentum $k^1_j  = 2\arctan (\Lambda_j^1)$, Eq. (\ref{kn-gen-Lambda}) 
for $n=1$, to the rapidity momentum 
$k_j$ of Ref. \cite{Taka}, such that $\Lambda_j^1 = \cot (k_j/2)$, is $k^1_j= \pi - k_j$.) 

The function $M_n (q_j)$ in Eq. (\ref{gen-Lambda}) is the $n$-band momentum distribution function associated with each
energy and momentum eigenstate. It is such that $M_n (q_j)=1$ and $M_n (q_j)=0$ for occupied and non-occupied $q_j$ values,
respectively. Such variables,
\begin{equation}
q_j = {2\pi\over L}I_j^n \, , \hspace{0.25cm} j=1,...,M_n^b \, ,
\label{qj}
\end{equation}
are the momentum values of a {\it $n$-band}. It is associated with the set of $M_n$ $n$-pair configurations with the
same $n$ value. 

On the one hand, the TBA $n=1$ band refers to the general BA spin band considered above. On the other hand, 
in the case of the TBA the $n$-pair configurations with $n>1$ spin-singlet pairs bound within them are also
associated with $n$-band sets of $M_n^b$ real momentum values, Eq. (\ref{qj}).
Here $M_n^b = M_n + M^h_n$ where the numbers $\{M_n\}$ of occupied momentum values in each 
such a $n$ band obey the sum rule $\sum_{n=1}^{\infty}n\,M_n = M_{\rm sp}$, Eq. (\ref{Nsingletpairs}).
The corresponding unoccupied values $\{M^h_n\}$ are uniquely defined by the spin $S$ and 
occupied values $\{M_n\}$ as follows \cite{Taka,CPC},
\begin{equation}
M^h_{n} = m^h_n\,L \, ; \hspace{0.25cm}
m^h_n = m_S+\sum_{n'=n+1}^{\infty}2(n'-n)\,m_{n'} \, .
\label{Mh}
\end{equation}
Moreover, the quantum numbers $I_j^n$ 
on the right-hand side of Eq. (\ref{qj}) are successive integers or half-odd integers according to the boundary conditions,
\begin{eqnarray}
I_j^n & = & 0,\pm 1,...,\pm {M_n^b -1\over 2} \, , \hspace{0.20cm}{\rm for}\hspace{0.20cm}M_n^b\hspace{0.2cm}{\rm odd} \, ,
\nonumber \\
& = & \pm 1/2,\pm 3/2,...,\pm {M_n^b -1\over 2} \, , \hspace{0.20cm}{\rm for}\hspace{0.20cm}M_n^b\hspace{0.2cm}{\rm even} \, ,
\label{Ijn}
\end{eqnarray}
respectively.

For each string of length $n$ there is thus a BA branch momentum $n$-band whose successive set of momentum values $q_j$, Eq. (\ref{qj}), 
have the usual separation, $q_{j+1}-q_{j}=2\pi/L$, and only occupancies zero and one.  
Often an index $\alpha = 1,...,M_n$ is used to label the subset of occupied quantum numbers $I_{\alpha}^n$ 
of an energy and momentum eigenstate \cite{Taka,CPC}. 

In the case of the TBA, we associate a {\it $n$-band pseudoparticle} with each of 
the $M_n$ $n$-band occupied momentum values \cite{CPC}. For $n>1$ the $n$-band pseudoparticles 
are specific to the TBA. On the one hand, the $M_n$ occupied $n$-band momentum values $q_j$ refer to their
translational degrees of freedom. They are associated with the center of mass motion of the
$M_n$ $n$-band pseudoparticles of momentum $q_j$. The corresponding $M^h_n$ 
unoccupied momentum values $q_j$ left over are associated with $M^h_n$ {\it $n$-band holes}.
Within a corresponding real-space lattice representation,
they interchange position with the $n$-band pseudoparticles under their center of mass motion.
On the other hand, the internal degrees of freedom of a
$n$-band pseudoparticle correspond to a single unbound spin-singlet pair for $n=1$ and to
a $n$-pair configuration with $n$ spin-singlet pairs bound within it for $n>1$. 

The $n$-band momentum distribution function $M_n (q_j)$
obeys the sum rule $\sum_{j =1}^{M_n^b} M_n (q_j) = M_n$. 
Each reduced subspace spanned by the set of energy and
momentum eigenstates with fixed spin $S$ and fixed number values $\{M_n\}$ 
has dimension ${M_n^b\choose M_n}$. It corresponds to the available different occupancy 
configurations of the $M_n$ $n$-band pseudoparticles over the $M_n^b$ momentum values.

The {\it exact momentum eigenvalues} have the simple form,
\begin{equation}
P = \pi + \sum_{n=1}^{\infty}\sum_{j=1}^{M_n^b}M_{n} (q_{j})\,q_j \, .
\label{P}
\end{equation}
This is consistent with the $n$-branch quantum numbers $q_j$, Eq. (\ref{qj}), playing the role of $n$-band pseudoparticle momentum 
values. 

There are sum rules for the number of $n$-band pseudoparticles 
that populate the $n=1,...,\infty$ bands of a LWS or non-LWS. Such sum rules
are related to those of spin-singlet pairs and density of spin-singlet pairs, Eqs. (\ref{Nsingletpairs0}) and (\ref{Nsingletpairs}).
Indeed, the latter sum rule implies that  $M_1 = M_{\rm sp} -  \sum_{n=2}^{\infty}n\,M_n$ and
thus that $M_1 = L(1-m_S)/2 -  \sum_{n=2}^{\infty}n\,M_n$. From the use of this relation in the number of 
pseudoparticles belonging to all $n=1,...,\infty$ bands, $M_{\rm ps}\equiv\sum_{n=1}^{\infty}M_n$, one confirms
that the following exact sum rules for $M_{\rm ps}$ and $m_{\rm ps}=M_{\rm ps}/L$ are obeyed,
\begin{equation}
M_{\rm ps} = \sum_{n=1}^{\infty}M_n = {1\over 2}(L-M^h_1) = m_{\rm ps}\,L \, ; \hspace{0.50cm}
m_{\rm ps} = \sum_{n=1}^{\infty}m_n = {1\over 2}(1-m^h_1) \, ,
\label{Nps}
\end{equation}
where the density $m^h_1=M^h_{1}/L$ refers to the number $M^h_{1}$ of $n=1$ band holes, Eq. (\ref{Mh}) for $n=1$. 

As a result of the TBA exact sum rule, Eq. (\ref{Nps}), the number of $n=1$ band holes $M^h_1$ and corresponding 
density $m^h_1$ play an important role in our study. They can be written in terms of the density of spin-singlet pairs
$m_{\rm sp}$, Eq. (\ref{Nsingletpairs}), and density of pseudoparticles $m_{\rm ps}$,
Eq. (\ref{Nps}), as follows,
\begin{equation}
M^h_1 = m^h_1\,L \, ; \hspace{0.50cm}
m^h_1 = m_S + 2(m_{\rm sp} - m_{\rm ps}) = m_S + 2(m_{\rm sp}^B - m_{\rm ps}^B) \, .
\label{mh1}
\end{equation}
The numbers $M_{\rm sp}^B$ of bound spin-singlet pairs and $M_{\rm ps}^B$ 
of $n>1$ band pseudoparticles within which they are bound and the corresponding
densities $m_{\rm sp}^B=M_{\rm sp}^B/L$ and $m_{\rm ps}^B=M_{\rm ps}^B/L$,
respectively, appearing in Eq. (\ref{mh1}) are given by,
\begin{eqnarray}
M_{\rm sp}^B & = & m_{\rm sp}^B\,L \, ; \hspace{0.25cm}
m_{\rm sp}^B = \sum_{n=2}^{\infty}n\,m_n = {1\over 2}(1-m_S) - m_1 \, ,
\nonumber \\
M_{\rm ps}^B & = & m_{\rm ps}^B\,L \, ; \hspace{0.25cm}
m_{\rm ps}^B = \sum_{n=2}^{\infty}m_n = {1\over 2}(1-m^h_1) - m_1 \, .
\label{NBspps}
\end{eqnarray}

As in Ref. \cite{CPC}, $m^{h,0}_{1}=M^{h,0}_{1}/L$ and $m_1^0=M_1^0/L$ denote corresponding densities of energy 
and momentum eigenstates with spin $S=0$. Those are given by $m^{h,0}_{1} = \sum_{n=2}^{\infty}2(n-1)m_{n}$
and $m_1^0 = 1/2 - \sum_2^{\infty}n\,m_n$, respectively. Hence,
$m^{h,0}_{1} = 2(m_{\rm sp}^B - m_{\rm ps}^B)$ and $m_1^0 =1/2 - m_{\rm sp}^B$. One then finds that
$m_{\rm sp}^B = {1\over 2}m^{h,0}_{1} + m_{\rm ps}^B$.
Similarly, $M^{h,0}_n = \sum_{n'=n+1}^{\infty}2(n'-n)\,M_{n'}$ and $m^{h,0}_n = \sum_{n'=n+1}^{\infty}2(n'-n)\,m_{n'}$.

The number $M_{\rm ps}$ in Eq. (\ref{Nps}) of pseudoparticles belonging to all $n=1,...,\infty$ bands
equals within the TBA that in Eq. (\ref{MpMhMbcomplex}) of $M_{\rm st} = M ^p + M_{\rm st}^B$
strings of all lengths $n=1,...,\infty$. Also the number $M_{\rm ps}^B$ in Eq. (\ref{NBspps}) of pseudoparticles 
of $n>1$ bands equals within the TBA that of $M_{\rm st}^B$ strings of length $n>1$. 
As discussed below in Sec. \ref{stringHTh0}, the unbinding of spin-singlet pairs by the collapse of narrow 
pairs is for a very large system and finite magnetic field the aberration from the ideal strings that
may have effects on the spin currents values.
Such processes are behind the inequalities $M_{\rm st}\geq M_{\rm ps}$ and $M_{\rm st}^B\leq M_{\rm ps}^B$
that apply to energy and momentum eigenstates described by groups of real and complex rapidities
within the general BA for a large system relative to those of the corresponding TBA states
in the TL. The equalities in these relations are reached when the string deformations of the former
states do not lead to the collapse of narrow pairs.

On the one hand, in the case of a LWS or non-LWS with $M_{\rm st}^B$ deformed strings of
length $n>1$ the corresponding independent $n$-pair configurations cannot be associated 
with $n$-band pseudoparticles carrying a real momentum $q_j$. 
On the other hand, the $M^p$ real rapidities of a LWS or non-LWS are both within the
general BA for a large system and the TBA in the TL associated with $M^p$ pseudoparticles whose internal
degrees of freedom refer to a single unbound spin-singlet pair. 

\section{General expressions for the spin stiffness at high temperature $T\rightarrow\infty$}
\label{stiffnessHT}

For $\vert S^z\vert \geq  1/2$, high temperature $T\rightarrow\infty$, and $L\rightarrow\infty$ 
the spin stiffness, Eq. (\ref{D-all-T-simp}), can in the TL be written as,
\begin{eqnarray}
D (T) & = & {(2S^z)^2\over 2 L T}{\sum_{S=\vert S^z\vert}^{L/2} \sum_{l_{\rm r}}
{\vert\langle\hat{J}^z (l_{\rm r},S)\rangle\vert^2\over (2S)^2}\over\sum_{S=\vert S^z\vert}^{L/2} 
\left\{{L\choose M_{\rm sp}}-{L\choose M_{\rm sp}-1}\right\}} \, ,
\label{D-Tlarge0}
\end{eqnarray}
where $M_{\rm sp}=L/2-S$ and $\sum_{l_{\rm r}}$ is the sum over the 
${\cal{N}}_{\rm singlet} (S) = {L\choose M_{\rm sp}}-{L\choose M_{\rm sp}-1}$ 
independent spin-singlet configurations of each fixed-$S$ subspace. Those are associated with the
${\cal{N}} (S)=(2S+1)\,{\cal{N}}_{\rm singlet} (S)$ energy and momentum eigenstates that span it.

The spin stiffness, Eq. (\ref{D-Tlarge0}), can alternatively be written as,
\begin{eqnarray}
D (T) & = & {(2S^z)^2\over 2 L T}{\sum_{S=\vert S^z\vert}^{L/2} \sum_{l_{\rm r}}
{\vert\langle\hat{J}^z (l_{\rm r},S)\rangle\vert^2\over (2S)^2}\over\sum_{S=\vert S^z\vert}^{L/2} 
\sum_{\{M_n\}_{m_S}}\,\prod_{n =1}^{\infty} {M_n^b\choose M_n}} \, ,
\label{D-Tlarge}
\end{eqnarray}
where the summation $\sum_{\{M_n\}_{m_S}}$ is over all $n=1,...,\infty$ band occupancies that refer to the same
number $M_{\rm sp}=L/2-S$ of spin-singlet pairs.
Provided that one uses on the right-hand side of Eq. (\ref{D-Tlarge}) the exact spin currents absolute values,
$\vert\langle\hat{J}^z (l_{\rm r},S)\rangle\vert$, this spin stiffness expression is rigorous for $\vert S^z\vert \geq  1/2$, 
$T\rightarrow\infty$, and $L\rightarrow\infty$. It is approximation free because
when written as ${\cal{N}}_{\rm singlet} (S) = \sum_{\{M_n\}_{m_S}}\,\prod_{n =1}^{\infty} {M_n^b\choose M_n}$
the number of independent spin-singlet configurations ${\cal{N}}_{\rm singlet} (S)$ 
in each fixed-$S$ subspace of dimension ${\cal{N}} (S) = (2S+1)\,{\cal{N}}_{\rm singlet} (S)$
has {\it exactly} the same value as when expressed in terms of the number $L$ of physical spins $1/2$ and
the number $M_{\rm sp}=L/2-S$ of spin-singlet pairs, ${\cal{N}}_{\rm singlet} (S) = {L\choose M_{\rm sp}}-{L\choose M_{\rm sp}-1}$
or equivalently ${\cal{N}}_{\rm singlet} (S) = {L\choose L/2-S}-{L\choose L/2-S-1}$. Indeed,
the TBA has been inherently constructed in Ref. \cite{Taka} to the dimensions ${\cal{N}} (S)=(2S+1)\,{\cal{N}}_{\rm singlet} (S)$ 
of {\it all} fixed-$S$ subspaces being exact in terms of the set of all $n$-bands occupancy configurations 
corresponding to a fixed number $M_{\rm sp}=L/2-S$ of spin-singlet pairs.

This is shown specifically in Appendix A of Ref. \cite{Taka} for LWSs for which the number
of unpaired spins $1/2$ with down-spin projection reads $2S=-2S^z$. Due to symmetry, that proof applies as well to
the non-LWSs in the fixed-$S$ subspaces. The off-diagonal generators that transform a $S>0$ LWS into 
its $2S$ tower states merely flip the spins of the $2S$ unpaired spins $1/2$ without changing the LWS
configurations of the $M_{\rm sp}$ spin-singlet pairs involving that state
$L-2S$ paired spins $1/2$.  

Within the general BA equations, Eq. (\ref{gen-Lambda-BA}), the spin current 
expectation values in Eq. (\ref{D-Tlarge0})  of energy and momentum eigenstates
described only by groups of real rapidities read,
\begin{equation}
\langle\hat{J}^z (l_{\rm r},S)\rangle = \sum_{\alpha} \,j^S (q_{\alpha}) \, ,
\label{J-S}
\end{equation}
both for large but finite chains and the TL. Here $q_{\alpha}$ 
denotes the corresponding occupied values of the BA spin band and the elementary currents 
$j^S (q_j)$ are given by,
\begin{eqnarray}
j^S (q_j) & = & - {2J\sin k_j\over 2\pi\sigma (k_j)}  \, ,
\nonumber \\
k_j & = & k (q_j) = 2\arctan (\Lambda_j) \, , \hspace{0.5cm} j = 1,...,M^b \, .
\label{j-k}
\end{eqnarray}
The distribution $2\pi\sigma (k_j)$ in the $j^S (q_j)$ expression obeys the following 
equation that within the TL can be transformed into an integral equation,
\begin{equation}
2\pi\sigma (k_j) = 1 - {1\over 2L\cos^2 (k_{j}/2)}\sum_{\alpha=1}^{M^p}{2\pi\sigma (k_{\alpha})\over  1 + \left({\tan (k_{j}/2) -
\tan (k_{\alpha}/2)\over 2}\right)^2} \, .
\label{sigma-IE-n1}
\end{equation} 
In this case the index $l_{\rm r}$ in Eq. (\ref{D-Tlarge0}) 
labels the $\sum_{l_{\rm r}}={\cal{N}}_{\rm singlet} (S) = {L\choose M_{\rm sp}}-{L\choose M_{\rm sp}-1}$ 
independent spin-singlet configurations of the $L-2S$ paired spins $1/2$ and corresponding
$M_{\rm sp}=L/2-S$ spin-singlet pairs associated with the
set of energy and momentum eigenstates that span each fixed-$S$ subspace.

In the general case of energy and momentum eigenstates described by groups of both by real 
and complex rapidities, there appear new types of contributions to 
the current operator expectation value expression, Eq. (\ref{J-S}). Such additional contributions
emerge from the strings of length $n>1$ associated with independent $n$-pair configurations with $n>1$ spin-singlet pairs 
bound within them. They can be computed from the use in the general BA equation, Eq. (\ref{gen-Lambda-BA}), of 
the suitable sets of specific complex rapidities of general form given in Eq. (\ref{Lambda-jnl}). 

Within the TBA, the spin currents $\langle\hat{J}^z (l_{\rm r},S)\rangle$ in Eq. (\ref{D-Tlarge})
of LWSs described by groups of real and complex rapidities can be written in the TL in terms of $n$-band 
pseudoparticle occupancies as follows \cite{CPC},
\begin{equation}
\langle\hat{J}^z (l_{\rm r},S)\rangle = \sum_{n=1}^{\infty}\sum_{j=1}^{M_n^b}\,M_n (q_j)\,\,j_n (q_j) \, .
\label{J-part}
\end{equation}
Here $l_{\rm r}$ labels the $\sum_{l_{\rm r}}={\cal{N}}_{\rm singlet} (S) = \sum_{\{M_n\}}\,\prod_{n =1}^{\infty} {M_n^b\choose M_n}$
independent spin-singlet configurations of the $L-2S$ paired spins $1/2$. They correspond to a
well-defined set of numbers $\{M_n\}$ of $n$-pair configurations associated with the energy and momentum 
eigenstates that span each fixed-$S$ subspace. The $n$-band elementary currents 
$j_n (q_j)$ in Eq. (\ref{J-part}) read \cite{CPC},
\begin{equation}
j_n (q_j) = - {2J\sin k^n (q_j)\over 2\pi\sigma^n (k^n (q_j))} 
\, , \hspace{0.20cm}{\rm where}\hspace{0.20cm}q_j \in [-q_n^b,q_n^b] \, ,
\label{jn-fn}
\end{equation}
where $q_n^b = \pi\,m_n^b$, the LWS rapidity functions $k^n (q_j)$ are obtainable 
from solution of the TBA equations, Eq. (\ref{Theta}), and within the TL the distribution 
$2\pi\sigma^n (k_j)$ is given by,
\begin{equation}
2\pi\sigma^n (k_j) \equiv 2\pi\sigma^n (k)\vert_{k=k_j} \, ; \hspace{0.5cm}
2\pi\sigma^n (k) = {\partial q^n (k)\over \partial k} \, .
\label{sigmsrhoderiv}
\end{equation}
Here $q^n (k)$ stands for the inverse function of the $n$-band rapidity momentum function $k^n (q)$. 

In Appendix \ref{ncurrentsG} it is found that
for LWSs for which $m_1^h \ll 1$ and $(1-m_1^h) \ll 1$ the elementary currents, Eq. (\ref{jn-fn}),
have the following exact limiting behaviors for the $n=1$ band,
\begin{eqnarray}
j_1 (q_j) & = & - J{\pi\over 2}\sin (q_j) \, , \hspace{0.20cm}{\rm for}\hspace{0.20cm}m_1^h \ll 1 
\nonumber \\
& = & - 2J\,\sin (q_j) \, , \hspace{0.20cm}{\rm for}\hspace{0.20cm}(1-m_1^h) \ll 1 \, .
\label{j1qj}
\end{eqnarray}
For the $n>1$ bands the corresponding exact limiting behaviors are,
\begin{eqnarray}
j_n (q_j) & = & - J{(n-1)\over 3n}(2\pi m_1^h)^2\sin \left({q_j\over m_n^b}\right)
\, , \hspace{0.20cm}{\rm for}\hspace{0.20cm}m_1^h \ll 1  
\nonumber \\
& = & - 2J\,\sin (q_j) \, , \hspace{0.20cm}{\rm for}\hspace{0.20cm}(1-m_1^h) \ll 1 \, .
\label{jS1n}
\end{eqnarray}
In addition, in that Appendix some of the exact behaviors useful for our studies of such elementary 
currents for a class of energy and momentum eigenstates whose currents absolute 
values reach largest values are reported. 

\section{The case of strictly zero magnetic-field}
\label{stiffnessHTh0only}

The general consensus is that the use of ideal strings for the energy and momentum eigenstates
described by groups of real and complex rapidities of the spin-1/2 $XXX$ Heisenberg chain
leads in the TL to exact results as long as either the temperature or the magnetic field are nonzero
\cite{Takahashi-03}. Concerning the spin stiffness, our results 
refer to $T\rightarrow \infty$, so that they are not affected in the TL by the finite-system string 
deformations. 

A technical difference between the cases $h=0$ and $h\neq 0$ is that 
for the former case of strictly zero magnetic-field there may occur deformations whose deviations $D_j^{n,l}$ from
the ideal string behavior may not occur in the strings themselves, Eq. (\ref{Lambda-jnl}). Hence at zero field
the problem is more complex in terms of the BA solution than for $h\neq 0$ and the use of the ideal strings 
in the BA equations to compute current operator expectation values of the corresponding $S^z=0$ energy 
and momentum eigenstates is often considered questionable, even in the TL. 

Fortunately, though, the current operator expectation values of these $S^z=0$ states, both
with spin $S=0$ and $S>0$, can be computed by a method that does not rely on the BA and TBA. It is then found that such 
expectation values exactly vanish \cite{CPC}. In the TL this applies both to energy and 
momentum eigenstates described by ideal and deformed strings of length $n>1$.

In order to briefly revisit that problem, we consider a class of spin current operator 
expectation values $\langle l_{\rm r} ,S,S^z\vert\hat{J}^z\vert l_{\rm r},S,S^{z}\rangle$
for energy and momentum eigenstates with arbitrary $S\geq 1/2$ and $S^{z}$ values
for which the following relation is exact \cite{CPC},
\begin{equation}
\langle  l_{\rm r} ,S,S^z\vert\hat{J}^z\vert  l_{\rm r},S,S^z\rangle = -{S^z\over S}
\langle  l_{\rm r} ,S,-S\vert\hat{J}^z\vert  l_{\rm r},S,-S\rangle \, , 
\label{currents-gen}
\end{equation}
where $S^z = -S + n_s$ and $n_s = 1,...,2S$.
This relation is obtained by combining the systematic use of the commutators given in Eq. (\ref{comm-currents}) 
with the state transformation laws $\hat{S}^{-}\vert l_{\rm r} ,S,0\rangle = 0$ and
$\hat{S}^{+}\vert l_{\rm r},0,0\rangle = \hat{S}^{-}\vert l_{\rm r},0,0\rangle = 0$,
which follow straight-forwardly from the corresponding spin $SU(2)$ symmetry operator algebra. 
The calculations to reach Eq. (\ref{currents-gen}) are relatively easy for non-LWSs whose generation 
from LWSs involves small $n_s=S-S^z$ values. As discussed in Ref. \cite{CPC},
they become lengthy as the $n_s$ value increases, but they remain straightforward. 
The exact relation, Eq. (\ref{currents-gen}), is behind the $T>0$ spin stiffness 
expression given in Eq. (\ref{D-all-T-simp}).

The form of the spin currents, Eq. (\ref{currents-gen}), confirms that the 
$S^z = 0$ expectation values $\langle  l_{\rm r} ,S,0\vert\hat{J}^z\vert  l_{\rm r},S,0\rangle$ indeed all vanish 
exactly for $S\geq 1/2$. The $S=S^z=0$ spin currents, $\langle l_{\rm r} ,0,0\vert\hat{J}^z\vert l_{\rm r},0,0\rangle$, 
are also found to vanish. They refer to energy and momentum eigenstates 
$\vert  l_{\rm r} ,0,0\rangle$ which are both LWSs and HWSs. 
It follows from Eq. (\ref{comm-currents}) that the current operator 
$\hat{J}^z$, Eq. (\ref{Jz-current}), may be expressed in terms 
of the commutator, $\hat{J}^z = {1\over 2}[\hat{J}^{+},\hat{S}^{-}]$.
Thus the spin currents $\langle l_{\rm r} ,0,0\vert\hat{J}^z\vert l_{\rm r},0,0\rangle$ can be written as,
$(\langle l_{\rm r} ,0,0\vert\hat{J}^{+}\hat{S}^{-}\vert l_{\rm r},0,0\rangle
- \langle l_{\rm r} ,0,0\vert\hat{S}^{-}\hat{J}^{+}\vert l_{\rm r},0,0\rangle)/2$.
That this expression vanishes is readily confirmed by applying the 
above state transformation laws. Hence all $S^z = 0$ spin currents 
$\langle  l_{\rm r} ,S,0\vert\hat{J}^z\vert  l_{\rm r},S,0\rangle$ vanish for $S\geq 0$.

The number and density of spin-singlet pairs reach their maximum values, $M_{\rm sp}=L/2$ and $m_{\rm sp}=1/2$, 
respectively, at $S=0$. Within both the general BA and the TBA, the $S=0$ absolute ground state has numbers values $M_{\rm sp}=L/2$,
$M_{\rm st}=M_{\rm ps}=M_1=M_{\rm sp}=L/2$ and thus $M_1^{h,0}=0$. For the TBA this implies that $M_n=0$ for $n>1$. 
For that ground state the spin/$n=1$ band is full. It has a symmetrical pseudoparticle compact momentum
occupancy. Hence such a state spin current exactly vanishes in the TL. 

Both $m_S=0$ and $m_1^{h,0}=0$ for such a ground state. In contrast, the remaining $S=0$ 
energy and momentum eigenstates may within the TBA have densities of $n=1$ band holes spanning the whole 
range, $m_1^{h,0} = \sum_{n=2}^{\infty}2(n-1)\,m_{n} \in [0,1]$. For each $n>1$ band pseudoparticle of momentum $q_j$ that populates 
such states, there are exactly $2(n-1)$ holes in the $n'=1$ band with momentum values $\{q_{j'}^h\}$ where $j' = 1,...,2(n-1)$.

What are the consequences of both in the case of the general BA and the TBA all $S^z = 0$ current expectation values 
$\langle  l_{\rm r} ,S,0\vert\hat{J}^z\vert  l_{\rm r},S,0\rangle$ vanishing for $S\geq 0$?
On the one hand, in the former case this implies the exact cancelling in the TL of the virtual elementary spin currents 
carried by the $M_{\rm st} = (L-M^h)/2$ independent spin-singlet pair configurations associated with strings of all lengths, $n=1,2,...,\infty$.
On the other hand, in the case of the TBA such an exact cancelling can be expressed 
in terms of the virtual elementary spin currents carried by the set of $M_n^{h,0} = \sum_{n'=n+1}^{\infty}2(n'-n)\,M_{n'}$ holes in 
the $n$-bands for which $M_n>0$. (The current contributions from $n$ bands for which $M_n=0$ vanish.)
Such set of $\sum_n M_n^{h,0}$ virtual elementary currents exactly cancel each other.

The $S^z=0$ energy and momentum eigenstates with spin $S>0$ have relative to the $S=0$ states an additional number 
of $2S$ holes in the spin/$n=1$ band. Within the TBA, all $n$ bands have an additional number of $2S$ holes.
An average number of $2S$ holes in that band and bands, respectively, now describe the translational degrees of freedom 
of the $2S$ unpaired spins $1/2$. Since $S$ of such unpaired spins have up-spin projection and the other $S$ unpaired spins have down-spin projection, 
the corresponding $S^z=0$ states with spin $S>0$ are non-LWSs. As confirmed in Sec. \ref{CurrentsUB}, 
for such states the additional virtual elementary current contributions from an average number of
$2S$ holes also cancel each other. (The remaining virtual current cancelling processes are 
similar to those of the $S=0$ energy and momentum eigenstates.)

Such a virtual current cancelation mechanism is encoded both in the general BA equation, Eq. (\ref{gen-Lambda-BA}),
and in the  $n = 1,...,\infty$ TBA equations, Eq. (\ref{gen-Lambda}), and corresponding general spin-current
expressions. However, for increasingly larger numbers of spin/$n=1$ band holes it is technically difficult to access from
direct solution of these equations. 

The problem can be explicitly solved in terms of such equations for the simplest case of the class of $S=0$ energy eigenstates with two holes
in the spin/$n=1$ band. Such states thus have one $n=2$-pair configuration described by one string of length two. (Within the TBA its 
two bound pairs refer to the internal degrees of freedom of one $n=2$ composite pseudoparticle.) 
This simplest case has been studied within the BA solution, as in Ref. \cite{Gu-07} for the present model, by use of the method
of Ref. \cite{Gu-07A} for the related large-on-site-repulsion half-filled 1D Hubbard model. (In this
paper the spin current operator, Eq. (\ref{Jz-current}), and its expectation values are given in units of $1/2$, 
which justifies that extra factor within the notation of Ref. \cite{Gu-07}.) One then explicitly finds that,
independently of the momentum values $q_j$ and $q_{j'}$ of the two holes, their virtual spin currents 
exactly cancel each other. 

As confirmed in the ensuing section, the virtual current mechanism also occurs for $\vert S^z\vert >0$
energy and momentum eigenstates. For such states it corresponds though to a partial cancellation \cite{CPC}.

\section{Useful inequalities and upper bounds on current absolute values}
\label{CurrentsUB}

The inequalities and corresponding current absolute values upper bounds introduced
in this section refer to the TBA. More general inequalities accounting for
the effects of the string deformations on the spin currents at finite magnetic field are
introduced below in Sec. \ref{stringHTh0}.

The spin-$1/2$ $XXX$ chain in a uniform vector potential $\Phi/L$ whose Hamiltonian is given
in Eq. (A2) of Ref. \cite{CPC} remains solvable by the BA. Within the TBA the LWSs momentum 
eigenvalues, $P = P (\Phi/L)$, have the general form,
\begin{equation}
P (\Phi/L) = P (0) + {L-\sum_{n=1}^{\infty}2n\,M_n\over L}\,\Phi  = P (0) + m_S\,\Phi = P (0) + 2S\,{\Phi\over L} \, .
\label{Peff}
\end{equation}
Here the $\Phi=0$ momentum eigenvalue $P (0)$ is given in Eq. (\ref{P}) and the sum rule $\sum_{n=1}^{\infty}2n\,M_n = L -2S$ involving the 
number $L -2S$ of paired physical spins $1/2$ has been used. (Such a sum rule
follows from that of the corresponding $M_{\rm sp} = L/2 -S$ spin-singlet pairs, Eq. (\ref{Nsingletpairs}).) Importantly, for large $L$
exactly the same exact momentum eigenvalues expression, $P (\Phi/L) = P (0) + 2S\,(\Phi/L)$,
is obtained by use of the BA accounting for deformed strings.

On the one hand, the expectation values of the current operator in the $\Phi\rightarrow 0$ LWSs, Eq. (\ref{J-part}), can be derived from the 
$\Phi/L$ dependence of the energy eigenvalues $E(\Phi/L)$ as $\langle \hat{J}^z\rangle = d E(\Phi/L)/d(\Phi/L)\vert_{\Phi=0}$ \cite{CPC}.
On the other hand, $d P(\Phi/L)/d(\Phi/L)\vert_{\Phi=0}$ gives the number of spin carriers that couple to the vector potential. 
The natural candidates are the model $L$ physical spins $1/2$. The form of the exact momentum eigenvalues, 
Eq. (\ref{Peff}), reveals that only the $2S$ unpaired spins $1/2$ contributing to the 
multiplet configurations couple to the vector potential $\Phi/L$. Since the $L-2S$ physical spins $1/2$ left over 
are those within the $M_{\rm sp} = L/2 - S$ {\it neutral} spin-singlet pairs, this exact result is physically appealing.

A second exact result is consistent with only the $2S$ unpaired physical spins $1/2$ coupling to the vector potential
also holding for non-LWSs. For simplicity, we consider that $L$ is even yet within the TL the same results are reached
for $L$ odd. For a general LWS carrying a spin current $\langle\hat{J}^z_{LWS} (l_{\rm r},S)\rangle$ all
$2S$ unpaired spins $1/2$ have up-spin projection. Let $S_{\sigma}$ be the number of
unpaired spins $1/2$ with spin projection $\sigma =\uparrow,\downarrow$ of a non-LWS such that
$\sum_{\sigma =\uparrow,\downarrow}S_{\sigma}=2S$. The exact relation, Eq. (\ref{currents-gen}),
can then be written simply as,
\begin{eqnarray}
\langle\hat{J}^z (l_{\rm r},S_{\uparrow},S_{\downarrow})\rangle & = & 
{(S_{\uparrow} - S_{\downarrow})\over 2S}\,\langle\hat{J}^z_{LWS} (l_{\rm r},S)\rangle 
\nonumber \\
& = & S_{\uparrow}\times j_{+1/2} + S_{\downarrow}\times j_{-1/2} \, ,
\label{currentsnLWS}
\end{eqnarray}
where
\begin{equation}
j_{\pm 1/2} = \pm {\langle\hat{J}^z_{LWS} (l_{\rm r},S)\rangle\over 2S} 
= \pm {1\over 2S}\sum_{n=1}^{\infty}\sum_{j=1}^{M_n^b}\,M_n (q_j)\,\,j_n (q_j) \, .
\label{jelementary}
\end{equation}

The exact relation, Eqs. (\ref{currentsnLWS}) and (\ref{jelementary}), 
confirms that only the $2S=S_{\uparrow} + S_{\downarrow}$ unpaired spins $1/2$
contribute to the spin currents. For each spin flip generated by application of the off-diagonal spin generator 
${\hat{S}}^{+}$ (and ${\hat{S}}^{-}$) onto an energy eigenstate with finite numbers $S_{\uparrow}$ 
and $S_{\downarrow}$, the spin current exactly changes by a {\it LWS current quantum} $2j_{-1/2}$ (and $2j_{+1/2}$.) 
Hence each unpaired spin $1/2$ with spin projection $\pm 1/2$ carries an elementary current $j_{\pm 1/2}$,
Eq. (\ref{jelementary}). For a LWS one has that $S_{\uparrow}=2S$ and $S_{\downarrow}=0$, so that
$\langle\hat{J}^z_{LWS} (l_{\rm r},S)\rangle = 2S\times j_{+1/2}$.

That only the $2S = m_S\,L$ unpaired physical spins $1/2$ couple to the vector potential
justifies the validity of the result of Ref. \cite{CPC} that all spin currents exactly vanish as $m_S\rightarrow 0$.
This exact result can be used to confirm that, as found in that reference, within the canonical-ensemble description 
at fixed value of $S^z$, in the TL, and for nonzero temperatures the spin stiffness $D (T)$, Eq. (\ref{D-all-T-simp}), 
vanishes as $m_S\rightarrow 0$. The main goal of this paper is to extend that result to the
grand-canonical-ensemble description for $T\rightarrow\infty$.

Relying on the exact relation, Eq. (\ref{currents-gen}), the spin stiffness expressions given in Eqs. (\ref{D-all-T-simp}), 
(\ref{D-Tlarge0}), and (\ref{D-Tlarge}) involve only spin current expectation values $\langle\hat{J}^z (l_{\rm r},S)\rangle$ 
of LWSs. It is thus useful to consider here the {\it LWS fixed-$S$ subspace} that is spanned by the
${\cal{N}}_{\rm singlet} (S)$ LWSs with a given spin $S$. It is a subspace of the
larger fixed-$S$ subspace spanned by all ${\cal{N}} (S) = (2S+1)\,{\cal{N}}_{\rm singlet} (S)$ energy and
momentum eigenstates with the same spin $S$. 

A LWS fixed-$S$ subspace can be further divided into smaller LWS reduced subspaces for which both the number $2S$
of unpaired physical spins $1/2$ and that of pseudoparticles $M_{\rm ps}$ are fixed. 
The $M_1^{h} = 2S + 2(M_{\rm sp}-M_{\rm ps}) = 2S, ...,2S + 2(M_{\rm sp}-1)$ value is thus also fixed. Hence
such subspaces refer to fixed values of the densities $m_S \in [0,1]$ and $m_1^h \in [m_S,1]$.

Each LWS fixed-$S$ subspace contains one {\it real-rapidity reduced subspace}. It is spanned by {\it real-rapidity LWSs}  
for which $m_1^{h}=m_S$. All $M_{\rm sp}=L/2-S$ spin-singlet pairs that populate such LWSs are unbound and thus
$M_{\rm ps}=M_1=M_{\rm sp}=L/2-S$ and $M_n=0$ for $n>1$. We denote its finite numbers by
$M_1=M_S^p$, $M_1^h=M_S^h$, and $M_1^b = M_S^b$ where,
\begin{eqnarray}
M_S^p & = & L/2 - S \, ; \hspace{0.50cm} M_S^h = 2S \, ,
\nonumber \\
M_S^b & = & M_S^p + M_S^h = L/2 + S \, .
\label{MS}
\end{eqnarray}
All remaining reduced subspaces of a LWS fixed-$S$ subspace are called {\it complex-rapidity reduced subspaces}.
Indeed those are spanned by {\it complex-rapidity LWSs} described by groups of both real and complex rapidities. Their
$m_1^{h}>m_S$ values belong to the range $m_1^{h} \in [m_S,1]$. 

We denote by $\vert\langle\hat{J}^z_{LWS}\rangle\vert_{T (m_S,m_1^h)}$ the largest current absolute 
value of each LWS reduced subspace of a given LWS fixed-$S$ subspace. It is of the general form,
\begin{equation}
\vert\langle\hat{J}^z_{LWS}\rangle\vert_{L(m_S,m_1^h)} =
{c_T\over L}2J\,2S\,M_{\rm ps} =
c_T 2JL\,m_S\,m_{\rm ps} \, .
\label{Jzmax}
\end{equation}
The coefficient $c_T$ in this expression obeys the inequality $c_T\leq\pi$. It is a function of the densities $m_S$ 
and $m_1^h$ with the following limiting behaviors,
\begin{eqnarray}
c_T & = & {\pi\over 2} \hspace{0.5cm}{\rm for}\hspace{0.2cm}m_S = m_1^h\rightarrow 0 \, ,
\nonumber \\
& = & 1 \hspace{0.5cm}{\rm for}\hspace{0.2cm}m_S = m_1^h\rightarrow 1 \, ,
\label{cTmSm1h}
\end{eqnarray}
and
\begin{eqnarray}
c_T & = & {\sin (\pi m_S)\over m_S} \hspace{0.5cm}{\rm for}\hspace{0.2cm}m_S \in [0,1/2]\hspace{0.2cm}
{\rm and}\hspace{0.2cm} m_1^h\rightarrow 1 \, ,
\nonumber \\
& = & {1\over m_S} \hspace{0.5cm}{\rm for}\hspace{0.2cm}m_S \in [1/2,1]\hspace{0.2cm}
{\rm and}\hspace{0.2cm} m_1^h\rightarrow 1 \, .
\label{cTmS1}
\end{eqnarray}
On the one hand, for $m_S\rightarrow 0$ and $m_1^h \in [0,1]$ it is an increasing function of
$m_1^h$ given by $c_T = \pi\,c_1$ where $c_1 = 1/2$ for $m_1^h\rightarrow 0$ and 
$c_1 = 1$ for $m_1^h\rightarrow 1$.  On the other hand, for $m_1^h=m_S$
it is a decreasing function of $m_1^h$ whose limiting values are given in Eq. (\ref{cTmSm1h}).

The LWSs spin currents result from processes that are simpler to be described in terms
of local spins $1/2$ occupancy configurations in the spin-$1/2$ $XXX$ chain lattice. 
Within these processes, each $2n$-site configuration of the $M_{\rm ps} = \sum_{n=1}^{\infty}M_n$ pseudoparticles 
that populate a LWS interchanges position under its motion along the lattice with such a state 
single-site $2S$ unpaired physical spins $1/2$. This justifies why the largest current absolute value of 
a LWS reduced subspace is proportional to $2S\times M_{\rm ps}$, as given in Eq. (\ref{Jzmax}). 
Consistently, LWSs for which $2S=0$ and/or $M_{\rm ps} = 0$ carry no spin current. 

The degrees of freedom of the $2S$ unpaired spins $1/2$ are distributed over different quantum numbers
of the exact BA solution. They are the physical spins $1/2$ whose spin is flipped by the spin $SU(2)$
symmetry algebra off-diagonal generators. The spin degrees of freedom of the $S_{\uparrow}$ and $S_{\downarrow}$ 
unpaired spins $1/2$ with up and down spin projection, respectively, determine the
spin $S=(S_{\uparrow}+S_{\downarrow})/2$ and spin projection $S^z = - (S_{\uparrow}-S_{\downarrow})/2$
of all energy eigenstates. Their translational degrees of freedom are described in each $n$-band
by its $M_n^h = 2S + \sum_{n'=n+1}^{\infty}2(n'-n)\,M_{n'}$ holes. Hence in terms of the
exact solution quantum numbers the above local processes that generate the spin currents refer to
the relative occupancy configurations of the $M_n$ pseudoparticles and corresponding $M_n^h$ holes
in each $n$ band for which $M_n>0$. Consistently, the LWSs spin currents $\langle\hat{J}^z_{LWS} (l_{\rm r},S)\rangle$ 
in the general spin current expression $\langle\hat{J}^z (l_{\rm r},S_{\uparrow},S_{\downarrow})\rangle = 
([S_{\uparrow} - S_{\downarrow}]/2S)\,\langle\hat{J}^z_{LWS} (l_{\rm r},S)\rangle$, Eq. (\ref{currentsnLWS}), 
can alternatively be expressed in terms 
of pseudoparticles, as given in Eq. (\ref{J-part}), or of $n$-band holes. Within the latter representation, they read
$\langle\hat{J}^z (l_{\rm r},S)\rangle = \sum_{n=1}^{\infty}\sum_{j=1}^{M_n^b}\,M_n^h (q_j)\,\,j_n^h (q_j)$
where $M_n^h (q_j) = 1 - M_n (q_j)$ and $j_n^h (q_j)=-j_n (q_j)$.

For non-LWSs, one can consider that $M_n^h = M_{n,\uparrow}^h + M_{n,\downarrow}^h$ where
$M_{n,\sigma}^h = S_{\sigma} + \sum_{n'=n+1}^{\infty}(n'-n)\,M_{n'}$ for $\sigma = \uparrow,\downarrow$.
The role of the additional number $\sum_{n'=n+1}^{\infty}2(n'-n)\,M_{n'}$ of holes in $M_n^h = 2S + \sum_{n'=n+1}^{\infty}2(n'-n)\,M_{n'}$
is to ensure that in each fixed-$S$ subspace dimension, ${\cal{N}} (S) = (2S+1)\,{\cal{N}}_{\rm singlet} (S)$, the factor
${\cal{N}}_{\rm singlet} (S)={L\choose M_{\rm sp}}-{L\choose M_{\rm sp}-1}$ where $M_{\rm sp}=L/2-S$
is exactly given by ${\cal{N}}_{\rm singlet} (S) = \sum_{\{M_n\}_{m_S}}\,\prod_{n =1}^{\infty} {M_n^b\choose M_n^h}$.

On the one hand, for the $S=0$ energy eigenstates considered in Sec. \ref{stiffnessHTh0only}, the number $M_n^h$ reads
$M_n^h = M_n^{h,0} = \sum_{n'=n+1}^{\infty}2(n'-n)\,M_{n'}$. In this case
the elementary currents carried by $\sum_{n'=n+1}^{\infty}(n'-n)\,M_{n'}$ $n$-band holes
exactly cancel those carried by the remaining $\sum_{n'=n+1}^{\infty}(n'-n)\,M_{n'}$ such holes.
On the other hand, for $S>0$ energy eigenstates for which $\sum_{n'=n+1}^{\infty}2(n'-n)\,M_{n'}>0$ there is a corresponding partial
elementary current cancellation. In that case out of the $M_n^h = 2S + \sum_{n'=n+1}^{\infty}2(n'-n)\,M_{n'}$ $n$-band holes
there is in average number $2S$ of such holes that describe the translational degrees of freedom of the
$2S$ unpaired spins $1/2$. Hence their elementary currents contribute to
the LWSs spin currents. The elementary currents carried by an average number 
$\sum_{n'=n+1}^{\infty}2(n'-n)\,M_{n'}$ of $n$-band holes cancel each other. In the case of LWSs, such a partial cancelling does
not occur in $n$-bands for which $M_n^h = 2S$. 

We denote by $\vert\langle\hat{J}^z_{LWS}\rangle\vert_{A(m_S,m_1^h)}$ 
the average current absolute value of each LWS reduced subspace.
It is given by,
\begin{equation}
\vert\langle\hat{J}^z_{LWS}\rangle\vert_{A(m_S,m_1^h)} =
{\sum_{l_{m_S,m_1^h}}\vert\langle\hat{J}^z (l_{m_S,m_1^h})\rangle\vert\over 
\sum_{\{M_n\}_{m_S,m_1^h}}\prod_{n =1}^{\infty} {M_n^b\choose M_n}} \, .
\label{JzAveDef}
\end{equation}
Here the sum $\sum_{l_{m_S,m_1^h}}$ runs over all $n=1,...,\infty$ band occupancy configurations
that generate the $\sum_{\{M_n\}_{m_S,m_1^h}}\prod_{n =1}^{\infty} {M_n^b\choose M_n}$ LWSs with the
same number $2S$ of unpaired physical spins $1/2$ and $M_{\rm ps}$ of pseudoparticles.
Hence the summation $\sum_{\{M_n\}_{m_S,m_1^h}}$ is over all sets of
$n$-band pseudoparticle numbers $\{M_n\}$ that obey both
the sum rules $\sum_{n=1}^{\infty}n\,M_n = {1\over 2}(L-2S)=M_{\rm sp}$, Eq. (\ref{Nsingletpairs}),
and $\sum_{n=1}^{\infty}M_n = {1\over 2}(L-M^h_1)=M_{\rm ps}$, Eq. (\ref{Nps}), respectively.

That each fixed-$S$ reduced subspace is spanned by energy eigenstates with exactly
the same number $M_{\rm ps}$ of pseudoparticles simplifies the
form of the average current absolute values, Eq. (\ref{JzAveDef}). In the TL they are related 
to the corresponding largest current absolute values, Eq. (\ref{Jzmax}), as follows,
\begin{eqnarray}
\vert\langle\hat{J}^z_{LWS}\rangle\vert_{A(m_S,m_1^h)} & = &
{c_A\over c_T}{\vert\langle\hat{J}^z_{LWS}\rangle\vert_{L(m_S,m_1^h)}\over \sqrt{2M_{\rm ps}}}
\nonumber \\
& = & {c_A\over L} J\,2S\,\sqrt{2M_{\rm ps}} \approx J\,m_S\,\sqrt{2M_{\rm ps}} \, .
\label{JzAve}
\end{eqnarray}
The coefficient $c_A$ reads here $c_A= 1$ for $(1-m_1^h)\ll 1$ and otherwise 
obeys the inequality $c_A\leq 1$, being of the order of unity. The factor $1/\sqrt{2M_{\rm ps}}$ that multiplies 
$\vert\langle\hat{J}^z_{LWS}\rangle\vert_{L(m_S,m_1^h)}$
stems from the LWSs that span the reduced subspace being generated by all 
possible occupancy configurations of the $M_{\rm ps}$ pseudoparticles.

In the case of the reduced subspace for which $M_{\rm ps}$ reaches its maximum value at fixed $S$,
the average current absolute value general form, Eq. (\ref{JzAve}), follows from the
calculations of Appendix \ref{currentUB2}. Its generalization to
the remaining reduced subspaces involves in the TL lengthy yet straightforward calculations. 
The precise value of the coefficient $c_A$ remains though an involved open problem.
Fortunately, the only related information needed for our studies is that $c_A$ is of the
order of the unity. 

At fixed spin $S$ the number $2S$ of unpaired physical spins $1/2$ that couple to a vector potential is fixed.
Hence the current absolute values are largest for LWSs for which these $2S$ unpaired spins $1/2$
have a larger number $M_{\rm ps}$ of $n$-band pseudoparticles to interchange position with. 

On the one hand, for a given LWS fixed-$S$ subspace the average current absolute value is thus 
smallest for its $M_{\rm ps}=1$ reduced subspace. For it the $M_{\rm sp}=L/2-S$ spin-singlet 
pairs are all bound within a single {\it gigantic} $n=M_{\rm sp}=L/2-S$ pair-configuration. 
The single pseudoparticle of the LWSs that span such a LWS reduced subspace has one of the $j=1,...,2S+1$ 
momentum values $q_j = 0,\pm {2\pi\over L},...,\pm {2\pi\over L}\,(S -1),\pm {2\pi\over L}\,S$.
For such LWSs the $M_{\rm sp}=L/2-S$ spin-singlet pairs involving the $L-2S$ paired spins $1/2$ 
reach the smallest dilution relative to the $2S$ unpaired spins $1/2$. The spin current of these LWSs, 
$\langle\hat{J}^z_{LWS} (l_{\rm r},S)\rangle = \langle\hat{J}^z_{LWS} (q_j,S)\rangle = - 2J\sin q_j$, results from the motion of the
single gigantic pseudoparticle relative to a number $2S$ of $n=L/2-S$ band holes. Those describe the translational degrees of 
freedom of the $2S$ unpaired physical spins $1/2$. 

On the other hand, both the largest current absolute $\vert\langle\hat{J}^z_{LWS}\rangle\vert_{T (m_S,m_1^h)}$, 
Eq. (\ref{Jzmax}), and the average current absolute value $\vert\langle\hat{J}^z_{LWS}\rangle\vert_{A(m_S,m_1^h)}$,
Eq. (\ref{JzAve}), reach their maximum values for the real-rapidity reduced subspace for which $M_{\rm ps}=M_1 = M_{\rm sp}=L/2-S$ 
and thus $M_n=0$ for $n>1$. Its average current absolute value, Eq. (\ref{JzAveDef}), can be written as,
\begin{equation}
\vert\langle\hat{J}^z_{LWS}\rangle\vert_{A(m_S,m_S)} =
{\sum_{l_{m_S}}\vert\langle\hat{J}^z (l_{m_S})\rangle\vert\over 
{M_S^b\choose M_S^p}} \, .
\label{JzAveReal}
\end{equation}
The sum $\sum_{l_{m_S}}$ in this expression runs over the set of 
$n=1$ band occupancy configurations that generate the ${M_S^b\choose M_S^p}$ LWSs
with the same spin $S$ whose $M_S^p$, $M_S^h$, and $M_S^b$ numbers are given in Eq. (\ref{MS}). 

That at fixed $m_S=2S/L$ the average current absolute value
$\vert\langle\hat{J}^z_{LWS}\rangle\vert_{A(m_S,m_1^h)} \approx J\,m_S\,\sqrt{2M_{\rm ps}}$
in Eq. (\ref{JzAve}) where $M_{\rm ps}=1,...,M_{\rm sp}$ 
reaches the largest value for the real-rapidity reduced subspace for which
$M_{\rm ps}=M_{\rm sp}$ plays a key role in our analysis. This implies that
in each LWS fixed-$S$ subspace the set of
average current absolute values, Eqs. (\ref{JzAveDef}) and (\ref{JzAve}),
for which $m_1^h>m_S$ obey the inequality,
\begin{equation}
\vert\langle\hat{J}^z_{LWS}\rangle\vert_{A(m_S,m_1^h)} <
\vert\langle\hat{J}^z_{LWS}\rangle\vert_{A(m_S,m_S)} \hspace{0.5cm}
{\rm for}\hspace{0.2cm}m_S < m_1^h < 1 \, .
\label{JzAveIneq}
\end{equation}
Here $\vert\langle\hat{J}^z_{LWS}\rangle\vert_{A(m_S,m_S)}$
is the corresponding real-rapidity reduced subspace
average current absolute value, Eq. (\ref{JzAveReal}).

We call $\vert\langle\hat{J}^z_{LWS}\rangle\vert_{A(m_S)}$ 
the average current absolute value of a LWS fixed-$S$ subspace.
It reads,
\begin{equation}
\vert\langle\hat{J}^z_{LWS}\rangle\vert_{A(m_S)} =
{\sum_{l_{\rm r}}\vert\langle\hat{J}^z (l_{\rm r},S)\rangle\vert\over {\cal{N}}_{\rm singlet} (S)} =
{\sum_{l_{\rm r}}\vert\langle\hat{J}^z (l_{\rm r},S)\rangle\vert\over
\sum_{\{M_n\}_{m_S}}\,\prod_{n =1}^{\infty} {M_n^b\choose M_n}} \, .
\label{meanAs}
\end{equation}
As in Eq. (\ref{D-Tlarge}), the sum $\sum_{l_{\rm r}}$ in this expression runs over all $n=1,...,\infty$ band occupancy configurations
that generate the ${\cal{N}}_{\rm singlet} (S)=\sum_{\{M_n\}_{m_S}}\,\prod_{n =1}^{\infty} {M_n^b\choose M_n}$ LWSs with the
same spin $S$. As in that equation, the summation $\sum_{\{M_n\}_{m_S}}$ is thus over all sets of
$n$-band pseudoparticle numbers $\{M_n\}$ that obey the sum rule 
$\sum_{n=1}^{\infty}n\,M_n =M_{\rm sp}=L/2-S$, Eq. (\ref{Nsingletpairs}).
This corresponds to the set of all energy eigenstates with the same number $M_{\rm sp}=L/2-S$ of spin-singlet pairs
and different numbers $M_{\rm ps}=1,...,M_{\rm sp}$ of pseudoparticles.

That the inequalities, Eq. (\ref{JzAveIneq}), are valid for {\it all} reduced subspaces of any
LWS fixed-$S$ subspace for which $m_1^h > m_S$ straightforwardly implies the 
validity of the following related inequality, 
\begin{equation}
\vert\langle\hat{J}^z_{LWS}\rangle\vert_{A(m_S,m_S)} \geq
\vert\langle\hat{J}^z_{LWS}\rangle\vert_{A(m_S)} \, .
\label{ineJJ}
\end{equation}
Since that validity refers to {\it all} $S>0$ values, it ensures the validity, within the TL, of the following important inequality
used below in the analysis of Sec. \ref{stiffnessHTh0},
\begin{equation}
{\sum_{S}\sum_{l_{m_S}}
{\vert\langle\hat{J}^z ({l_{m_S}})\rangle\vert^2\over (2S)^2} \over\sum_{S} {M_S^b\choose M_S^p}} \geq 
{\sum_{S} \sum_{l_{\rm r}}{\vert\langle\hat{J}^z (l_{\rm r},S)\rangle\vert^2\over (2S)^2} \over\sum_{S} 
\sum_{\{M_n\}_{m_S}}\,\prod_{n =1}^{\infty} {M_n^b\choose M_n}} \, .
\label{ineq}
\end{equation}
Before presenting such an analysis, a more general inequality accounting for the effects of the string deformations
is introduced in the ensuing section.

\section{The effects of the string deformations on the spin currents at finite magnetic field}
\label{stringHTh0}

At finite magnetic field only the deviations $D_j^{n,l}$ that occur in the strings themselves, Eq. (\ref{Lambda-jnl}),
may have effects in the TL on the spin currents and other quantities.
The set of these complex rapidities with the same real part of form 
$\Lambda_j^{n,l} = \Lambda_j^n + i (n+1-2l) + D_j^{n,l}$
remain being labelled by the quantum numbers 
$n=1,...,\infty$ and $l = 1,...,n$ that refer to the number of bound spin-singlet pairs 
and each of these pairs, respectively. Physically, this means that, as in the case of 
an ideal string, the distorted string associated with that set of complex rapidities 
also describes an independent configuration within which
$n=1,...,\infty$ spin-singlet pairs are bound.

The set of TBA complex rapidities with the same real part, Eq. (\ref{Lambda-jnl-ideal}), 
obey the symmetry relation $\Lambda_j^{n,l} = (\Lambda_j^{n,n+1-l})^*$. The 
two complex rapidities $\Lambda_j^{n,l}$ and $\Lambda_j^{n,l'}$ associated with
two spin-singlet pairs labelled by the quantum numbers $l$ and $l'=n+1-l$, respectively, being related as
$\Lambda_j^{n,l}=(\Lambda_j^{n,l'})^*$ for $l = 1,...,n$ is actually a necessary condition for the binding of the
$l = 1,...,n$ spin-singlet pairs within the $n$-pair configuration.

Importantly and due to self-conjugacy, the deviations $D_j^{n,l} = R_j^{n,l} + i \delta_j^{n,l}$
in Eq. (\ref{Lambda-jnl}) for the set of complex rapidities with the same real
part associated with a distorted string
are also such that $D_j^{n,l} = (D_j^{n,n+1-l})^*$. This reveals that
the symmetry $\Lambda_j^{n,l} = (\Lambda_j^{n,n+1-l})^*$ prevails under string deformations.
This ensures that as for the ideal strings, the imaginary parts of the $n$ real rapidities with the
same real part associated with deformed strings also describe the
binding within the corresponding $n$-pair configurations of $l = 1,...,n$ spin-singlet pairs.

The V-strings deformations \cite{Takahashi-03} have in the TL and finite magnetic field 
no effects on the spin currents. At finite magnetic field the EKS-strings {\it collapse of narrow pairs}, described below within our representation
in terms spin-singlet pair unbinding processes, is in the TL the only aberration from the ideal strings 
\cite{Takahashi-03} that may have effects on the spin currents. This refers only to
the currents of $\vert S^z\vert >0$ energy and momentum eigenstates described by groups of real and complex rapidities. 
Here we identify such effects and justify why in the TL they have no impact whatsoever in the
high-temperature stiffness upper bounds introduced in the ensuing section.

The general consensus is that the use ideal strings for
energy and momentum eigenstates described by groups of real and complex rapidities
leads in the TL to exact results as long as either the temperature or 
the magnetic field are nonzero \cite{Takahashi-03}. Consistently, 
although the collapse of narrow pairs is indeed found to enhance the spin currents 
absolute values of a few states, it does no change in the TL the stiffness upper bounds
used in our study.

The string deviations from the TBA ideal strings do not change the value of 
the number of spin-singlet pairs. Hence their density is also exactly
given by $m_{\rm sp} = (1-m_S)/2$ for the corresponding LWSs and non-LWSs.
Narrow pairs refer to a string deformation originated by a deviation $D_j^{n,l}$ 
that renders the separation between two rapidities
$\Lambda_j^{n,l}$ and $\Lambda_j^{n,l+1}$ in the imaginary direction less than i.
Such a separation may become narrower and eventually merge and split back onto 
the horizontal axis \cite{Takahashi-03}. Such a process is what is called the collapse of a narrow pair. 

Within our representation in terms of the model physical spins $1/2$,
it then refers to an elementary process that leads to the unbinding of two spin-singlet pairs. On the one hand, for the
set of $n>2$ complex rapidities with the same real part associated with $n$ bound pairs, it leads to the partition of the
corresponding $n$-pair configuration into a $n'$-pair configuration where $n'=n-2$. The latter
is described by a smaller number $n'=n-2$ of complex rapidities with the same
real part in a string of smaller length $n'=n-2$. The process also generates two 
unbound spin-singlet pairs described by real rapidities. On the
other hand, for $n=2$ complex rapidities with the same real part it leads in turn
to the unbinding of the two spin-singlet pairs of the corresponding $n=2$ pair configuration.
This gives rise solely to the two unbound spin-singlet pairs described by real rapidities. 

Hence the collapse of a narrow pair is a process that causes an increase in the value of 
the number of strings of all lengths, $M_{\rm st} = M^p + M_{\rm st}^B$,
Eq. (\ref{MpMhMbcomplex}). It does not change though that of spin-singlet pairs, $M_{\rm sp} = L/2-S$. 
Specifically, it always leads to a positive deviation $\delta M^p=2$ in the value of the
number of spin-band pseudoparticles and corresponding unbound spin-singlet pairs. Moreover,
it gives rise to a negative deviation $\delta M_{\rm sp}^B = -2$ in the value of the number of bound spin-singlet pairs.
There is as well either an additional negative deviation $\delta M_{\rm st}^B=-1$ or no deviation $\delta M_{\rm st}^B=0$ in the number $M_{\rm st}^B$
of independent configurations with bound spin-singlet pairs within them. This 
depends on whether the deformed $n$-pair configuration that
suffers the collapse of a narrow pair has $n=2$ or $n>2$
spin-singlet pairs bound within it, respectively. 

We denote by $\vert\langle\hat{J}^z_{LWS}\rangle\vert_{A_D (m_S)}$ 
the average current absolute value of the LWS fixed-$S$ subspace spanned
by energy eigenstates for which some of the complex strings are deformed.
It is given by,
\begin{equation}
\vert\langle\hat{J}^z_{LWS}\rangle\vert_{A_D (m_S)} =
{\sum_{l_{\rm r_D}}\vert\langle\hat{J}^z (l_{\rm r_D},S)\rangle\vert\over {\cal{N}}_{\rm singlet} (S)} \, .
\label{meanAsD}
\end{equation}
The sum $\sum_{l_{\rm r_D}}$ in this expression runs over all $L-2S$ paired physical spins $1/2$ occupancy configurations
that generate the ${\cal{N}}_{\rm singlet} (S)$ LWSs with the
same spin $S$ and thus the same number $M_{\rm sp}=L/2-S$ of spin-singlet pairs.

As given in Eq. (\ref{ineJJ}), within the TBA the average of the current absolute values is largest in the fixed-$S$ subspaces
spanned by energy and momentum eigenstates described only by groups of real rapidities. Such an average is
larger than that in the fixed-$S$ subspaces spanned by all energy and momentum eigenstates of spin $S$.
The main point is that a larger fraction of unbound spin-singlet pairs relative to
bound spin-singlet pairs at the fixed number $M_{\rm sp} = L/2-S$ of
such pairs tends to enhance the spin current absolute values. 

A generalization of the inequality, Eq. (\ref{ineJJ}), which accounts for the effects of
the collapse of narrow pairs and thus of spin-singlet pair unbinding processes, 
involves the average current absolute value, Eq. (\ref{meanAsD}), and reads, 
\begin{equation}
\vert\langle\hat{J}^z_{LWS}\rangle\vert_{A(m_S,m_S)} \geq
\vert\langle\hat{J}^z_{LWS}\rangle\vert_{A_D (m_S)} \geq
\vert\langle\hat{J}^z_{LWS}\rangle\vert_{A(m_S)} \, .
\label{ineJJD}
\end{equation}

On the one hand, the validity of the inequality $\vert\langle\hat{J}^z_{LWS}\rangle\vert_{A(m_S,m_S)}\geq 
\vert\langle\hat{J}^z_{LWS}\rangle\vert_{A_D (m_S)}$ in this equation follows from the energy 
and momentum eigenstates described by real rapidities having no strings of length $n>1$ and thus being
string-deformation free. This is because all their $M_{\rm sp} = L/2-S$ spin-singlet pairs
are unbound. The binding of spin-singlet pairs within $n$-pair configurations for which $n>2$ in states
with groups of real and complex rapidities lessens the  
current absolute values. The unbinding of spin-singlet pairs under string deformations 
only partially neutralizes this effect. Indeed, it does not refer to all spin-singlet pairs bound within 
$n$-pair configurations for which $n>2$. In contrast, for the energy and momentum eigenstates described by real 
rapidities {\it all} $M_{\rm sp}=L/2 -S$ spin-singlet pairs are unbound. 

On the other hand, the inequality $\vert\langle\hat{J}^z_{LWS}\rangle\vert_{A_D (m_S)}\geq 
\vert\langle\hat{J}^z_{LWS}\rangle\vert_{A(m_S)}$ in Eq. (\ref{ineJJD}) is valid because
the collapse of narrow pairs caused by complex rapidity string
deformations may unbind some spin-singlet pairs. This effect tends to enhance
the average of the current absolute values in the fixed-$S$ subspaces 
whose strings of some states are deformed. This effect is though very small in the TL.
Indeed most string deformations involve small variations
in the string fine structure that do not lead to the collapse of narrow pairs and
in the TL have no effects on the spin currents absolute values.

Since the inequalities in Eq. (\ref{ineJJD}) are valid for all $S>0$ values,
the following important inequality, which is an extension of that given in Eq. (\ref{ineq}), holds,
\begin{equation}
{\sum_{S}\sum_{l_{m_S}}
{\vert\langle\hat{J}^z ({l_{m_S}})\rangle\vert^2\over (2S)^2} \over\sum_{S} {M_S^b\choose M_S^p}} \geq 
{\sum_{S} \sum_{l_{\rm r_D}}{\vert\langle\hat{J}^z (l_{\rm r_D},S)\rangle\vert^2\over (2S)^2} \over\sum_{S} 
{\cal{N}}_{\rm singlet} (S)} \, .
\label{ineqD}
\end{equation}

\section{High-temperature stiffness upper bounds within the thermodynamic limit}
\label{stiffnessHTh0}

The high-temperature stiffness upper bounds introduced in this section rely
on replacing averages of the spin current absolute values in the full LWS spin-$S$ subspaces
by those in the corresponding smaller LWS real-rapidity reduced subspaces. It follows from the inequalities,
Eqs. (\ref{ineq}) and (\ref{ineqD}), that our final results are independent 
from the use in the TL of ideal or deformed strings for the states described by groups of
real and complex rapidities. 

For simplicity, we use the number notation in Eq. (\ref{MS}), within which
$M_S^p (q_j) = M_1 (q_j)$, $q_j = (2\pi/L)\,I_{j}\in [-q^b,q^b]$, $I_{j}=I_{j}^1$,
and $q^b=q_1^b=\pi (M_{S}^b-1)/L$. Each LWS real-rapidity reduced subspace is then 
spanned by ${M_S^b\choose M_S^p}$ energy and momentum eigenstates with the same $S$ value.

A first spin stiffness upper bound, $D_{u1} (T)\geq D(T)$, is derived 
from the direct use in the high-temperature stiffness expression, Eq. (\ref{D-Tlarge}), 
of the inequalities in Eqs. (\ref{ineq}) and (\ref{ineqD}). This leads to,
\begin{equation}
D_{u1} (T) = {(2S^z)^2\over 2 L T}{\sum_{S=\vert S^z\vert}^{L/2}\sum_{l_{m_S}}
{\vert\langle\hat{J}^z (l_{m_S})\rangle\vert^2\over (2S)^2}\over\sum_{S=\vert S^z\vert}^{L/2} {M_S^b\choose M_S^p}} \, .
\label{D-T-u1}
\end{equation}
The sums $\sum_{l_S }$ in this expression run over the real-rapidity LWSs whose number is
${M_S^b\choose M_S^p}$ that span each LWS real-rapidity reduced
subspace. The spin currents $\langle\hat{J}^z (l_{m_S})\rangle$ are given by,
\begin{equation}
\langle\hat{J}^z (l_{m_S})\rangle=\sum_{j =1}^{M_S^b} M_S^p (q_j)\,j_1^S (q_j) \, ,
\label{J-partS}
\end{equation}
where $\sum_{j =1}^{M_S^b} M_S^p (q_j) = M_S^p$.
The elementary current $j_1^S (q_j)$ in this expression is that in
Eq. (\ref{j1Sgen}) of Appendix \ref{ncurrentsR}.
It reads $j_1^S (q_j) = j_1 (q_j)$ for $q_j\in [-q^b,q^b]$ and $M_1=M_{\rm ps}=M_{\rm sp}$ where 
$j_1 (q_j)$ is the elementary current, Eq. (\ref{jn-fn}) for $n=1$.

For the present real-rapidity LWSs one has that
$m_1^h = m_S$. Hence the limits given in Eq. (\ref{j1qj}) apply. 
The elementary current $j_1^S (q_j)$ changes thus from $j_1^S (q_j) = -{\pi\over 2}J\sin q_j$ for 
$q_j\in [-\pi/2,\pi/2]$ as $m_S\rightarrow 0$ to $j_1^S (q_j) = -2J\sin q_j$ for $q_j\in [-\pi,\pi]$ 
as $m_S\rightarrow 1$. It can be written as $j_1^S (q_j) = - j_1^S\,s_1^S (q_j)$ where
$\vert s_1^S (q_j)\vert\leq 1$ for $q_j \in \left[-{\pi\over 2}(1-m_S),{\pi\over 2}(1-m_S)\right]$.
As justified in Appendix \ref{ncurrentsR}, the elementary current coefficient $j_1^S>0$
in that expression reaches its largest value $j_1^S = 2J$ for the
whole $m_S \in [0,1]$ range for $m_S\rightarrow 1$. Moreover, in that
Appendix it is found that the replacement in $j_1^S (q_j) = - j_1^S\,s_1^S (q_j)$
of $j_1^S$ and $s_1^S (q_j)$ by $2J$ and $\sin q_j$, respectively, ensures that
$\vert\sum_{j =1}^{M_S^b} M_S^p (q_j)\,2J\sin q_j\vert\geq
\vert\sum_{j =1}^{M_S^b} M_S^p (q_j)\,j_1^S (q_j)\vert$ for all real-rapidity LWSs 
and the whole $m_S\in [0,1]$ interval. This thus implies the validity of the following inequality, 
\begin{equation}
\sum_{S=\vert S^z\vert}^{L/2}\sum_{l_{m_S}}
{J_{*}^2 (l_{m_S})\over (2S)^2} \geq
\sum_{S=\vert S^z\vert}^{L/2}\sum_{l_{m_S}}
{\vert\langle\hat{J}^z (l_{m_S})\rangle\vert^2\over (2S)^2} \, ,
\label{ineq2}
\end{equation}
where $J_{*} (l_{m_S}) = -\sum_{j =1}^{M_S^b} M_S^p (q_j)\,2J\sin q_j$.

Our second stiffness upper-bound, $D_{u2}(T)\geq D(T)$, 
is thus obtained by replacing in Eq. (\ref{D-T-u1})
the factor on the right-hand side of Eq. (\ref{ineq2}) by that on its left-hand side.
This accounts for replacing the exact elementary spin current $j_1^S (q_j)$ 
by a upper-bound elementary spin current given by,
\begin{equation}
j (q_j) = -2J\,\sin q_j \, .
\label{jn-fn0}
\end{equation}

Under this replacement, the sum $\sum_{l_{\rm r_S}}$ in Eq. (\ref{D-T-u1}) can be performed. Such a sum
is carried out in Appendix \ref{currentUB2}, with the result,
\begin{eqnarray}
D_{u2}(T)  & = & \frac{\sum_{S}^-  
\frac{J^2 (S^z)^2}{L T\,S^2}\left(M_{\rm sp} + 2S + \frac{\sin(2\pi S/L)}{\sin(2\pi/L)}\right) {M_{\rm sp} + 2(S-1) \choose M_{\rm sp} - 1}}
{\sum_{S} {M_{\rm sp} + 2S\choose M_{\rm sp}}} 
\nonumber \\
& = & \frac{\sum_{S}^-  
\frac{J^2 (S^z)^2}{L T\,S^2}\left(L/2+S+ \frac{\sin(2\pi S/L)}{\sin(2\pi/L)}\right) {L/2+S - 2 \choose L/2 -S - 1}}
{\sum_{S} {L/2+S\choose L/2-S}} \, ,
\label{bound}
\end{eqnarray}
for $T\rightarrow\infty$. Here the summations refer to $\sum_{S}^-=\sum_{S=|S^z|}^{L/2-1} $ and $\sum_{S}=\sum_{S=|S^z|}^{L/2}$, 
respectively, and for simplicity we have chosen $L$ to be even so that $S^z$ and $S$ are integers. (In the present TL this reaches
again the same final results as for $L$ odd.)

The following behaviors of the spin stiffness upper bound $D_{u2}(T)$, Eq. (\ref{bound}), corresponding to $m\ll 1$ and $(1-m)\ll 1$ are 
derived in Appendix \ref{currentUB2},
\begin{eqnarray}
D_{u2}(T) & = & \frac{J^2\,c_{u2}}{2T}\,m^2 \approx  \frac{J^2}{2T}\,m^2 \, , \hspace{0.25cm}{\rm for}\hspace{0.20cm}m \ll 1 \, ,
\nonumber \\
& = & \frac{J^2}{2T}\, (1-m)  \, , \hspace{0.25cm}{\rm for}\hspace{0.20cm}(1-m)\ll 1 \, ,
\label{value-D-m0-1}
\end{eqnarray}
respectively, where,
\begin{equation}
c_{u2} =  \frac{9}{4}(\sqrt{5} - 2)\left(\frac{5}{3} + \frac{\sqrt{3}}{2\pi}\right) \approx 1.032 \, .
\label{cu2}
\end{equation}

Finally, we emphasize that our $T\rightarrow\infty$ upper bound, Eq. (\ref{bound}), has been inherently constructed to the
exact $T\rightarrow\infty$ stiffness reading,
\begin{equation}
D(T) = D_{u2}(T) = {J^2\over 2T}\, (1-m) \, ,
\label{Dreal1}
\end{equation}
for $(1-m)\ll 1$ and, 
\begin{equation}
D(T) =  {J^2\,c^2\over 2T}\,m^2 \, ,
\label{Dreal2}
\end{equation}
for $m\ll 1$. Here $c$ is a $m$ and $T$ independent coefficient, $c\approx 1$ such that $c^2<c_{u2}$ .

The calculations of Appendix \ref{currentUB2} that reached the expressions in Eqs. (\ref{Dreal1}) and 
(\ref{Dreal2}) correspond in these two limits to average current absolute values of the form
$\vert\langle\hat{J}^z_{LWS}\rangle\vert_{A(m_S,m_S)} = c\,J\,m_S\,\sqrt{2M_{\rm ps}} =
c\,J\,m_S\,\sqrt{L-2S}$ where $c=1$ for $(1-m_S)\ll 1$ and
$c\approx 1$ for $m_S\ll 1$, consistently with Eq. (\ref{JzAve}) for $m_1^h = m_S$ where
$m_S=m$ for LWSs.

\section{Concluding remarks}
\label{concluding}

The upper bound on high-temperature spin stiffness derived in this paper, Eqs. (\ref{bound})-(\ref{cu2}), 
vanishes as $m^2$ in the $m\rightarrow 0$ limit and is independent of the system size $L$. This ensures that the spin stiffness vanishes 
within the {\it grand-canonical ensemble} as $h\rightarrow 0$ for high temperature $T\rightarrow\infty$ in the TL. 
We believe that our result is exact in these limits. As discussed in the following, the possibility of the 
absence of ballistic spin transport for the whole finite-temperature range $T>0$ within the grand-canonical ensemble 
in the limit of zero magnetic field remains though an interesting unsolved problem.

Concerning the relation of our results to previous results on the spin stiffness of the spin-$1/2$ $XXX$ chain,
the upper bound of Ref. \cite{CPC} is valid for the whole temperature range $T>0$ and vanishes as $m^2\,L$ in the $m\rightarrow 0$
limit. This latter behavior reveals that within the canonical ensemble the model spin stiffness vanishes as $m\rightarrow 0$ 
for finite temperature within the TL. However and as mentioned above, it leaves out, marginally, the grand canonical ensemble 
in which $\langle m^2\rangle = {\cal{O}}(1/L)$. The large overestimate of the current absolute values used in deriving the stiffness 
upper bound of that reference, whose limiting values are given in Eq. (\ref{value-allT-D-m0-1}), leads for high temperature 
to an extra factor of the order ${\cal{O}} ((1-m)L)$ relative to our upper bound, Eq. (\ref{bound}). This refers to
an overestimate of the method used in Ref. \cite{CPC} that has ignored the factor $1/\sqrt{2M_{\rm ps}}=1/\sqrt{(1-m_S)L}$ 
in the corresponding spin current average value, first expression of Eq. (\ref{JzAve}) for $m_1^h = m_S$ where $m_S=m$ for LWSs.

We note that our result on vanishing spin stiffness as $h\rightarrow 0$ in the TL 
crucially depends on the existence of a global $SU(2)$ symmetry where the current 
under comsideration is a part of the symmetry operator algebra. We thus expect that our result should 
be extendable to other integrable models with similar one or several global $SU(2)$ symmetries, 
such as {\it e.g.} the fermionic 1D Hubbard model. 

In conclusion, in this paper we addressed the important fundamental and highly debated question
on the possibility of ballistic spin transport within the grand-canonical 
ensemble for  $h\rightarrow 0$ in what is arguably one of the simplest strongly correlated quantum many-body system, the 
spin-$1/2$ $XXX$ chain. Our main result is the strong evidence of
lack of such a ballistic transport within the grand-canonical ensemble as $h\rightarrow 0$ in the TL 
at high temperature $T\rightarrow\infty$.

Our results thus imply that the spin-$1/2$ $XXX$ Heisenberg chain exhibits at infinite temperature 
anomalous sub-ballistic spin transport. This is consistent with the studies of Ref. \cite{Znidaric-11} 
that rely on a nonequilibrium open system approach. 
    
Combination of the result of Ref. \cite{CPC} that within the canonical ensemble the spin stiffness 
vanishes in the $m\rightarrow 0$ limit at all nonzero temperatures with the absence of phase transitions 
in the spin-$1/2$ $XXX$ chain at $T>0$, could be an indication of the lack of ballistic spin transport
for the whole nonzero temperature range, $T>0$, also within the grand-canonical ensemble. This remains
though an interesting open problem that deserves further studies.

Last but not least, our method uses a representation in terms of configurations of the 
$L$ physical spins $1/2$ that is more controllable than most numerical studies on the occurrence or lack of ballistic 
spin transport in the spin-1/2 XXX chain. Moreover, such a representation provides useful physical information on the 
microscopic processes involving the elementary currents carried by spin/$n=1$ band holes and $n$-pair configurations
with $n>1$ spin-singlet pairs bound within them that control the very complex problem under investigation. That information may 
play a valuable role in future studies of the present problem for the whole nonzero temperature range, $T>0$.
         
\acknowledgements
We thank David K. Campbell, Pedro. D. Sacramento, and Xenophon Zotos for discussions. J. M. P. C. thanks the support by 
Portuguese FCT through the Grant UID/FIS/04650/2013. T. P. acknowledges support by the Grants of Slovenian Research Agency (ARRS)
P1-0044, N1-0025, and J1-5439.

            
\appendix

\section{Spin-band and $n$-band elementary currents}
\label{ncurrents}

The main aim of this Appendix is to derive the elementary current $j_n (q_j)$ expressions, Eqs. (\ref{j1qj}) and (\ref{jS1n}),
and to justify the validity of the inequality, Eq. (\ref{ineq2}). To achieve such goals, 
expressions for the elementary currents $j_n (q_j)$, Eq. (\ref{jn-fn}), of classes
of LWSs that include those whose absolute values of the current
$\langle\hat{J}^z (l_{\rm r},S)\rangle = \sum_{n=1}^{\infty}\sum_{j=1}^{M_n^b}\,M_n (q_j)\,\,j_n (q_j)$, Eq. (\ref{J-part}), is larger
are derived. In the case of LWS fixed-$S$ real-rapidity reduced subspaces 
considered in Sec. \ref{stiffnessHTh0}, this refers to the elementary currents $j^S (q_j)$,
Eq. (\ref{J-S}), in the current expression, Eq. (\ref{j-k}). 

\subsection{$n$-band elementary currents for classes of LWSs described by groups of real and complex rapidities}
\label{ncurrentsG}

The goal of this Appendix section is to justify the validity of the elementary current $j_n (q_j)$ expressions, 
Eqs. (\ref{j1qj}) and (\ref{jS1n}).
It is straightforward to confirm from manipulations of the TBA equations, Eq. (\ref{gen-Lambda}),
LWS spin current expression, Eq. (\ref{J-part}), and corresponding $n$-band elementary
current expression, Eq. (\ref{jn-fn}), that the class of LWSs that reach the largest current absolute values have 
asymmetrical compact hole or pseudoparticle $n$-band distributions.
Here we consider the larger class of LWSs with compact hole or pseudoparticle $n$-band distributions
in the TL of the general form, 
\begin{eqnarray}
M_n^0 (q_j) & = & \Theta (q_j+\pi m_n^b)\Theta (q_{h\,n}^--q_j) 
\nonumber \\
& + & \Theta (q_j-q_{h\,n}^+)\Theta (\pi m_n^b-q_j) \, , \hspace{0.20cm}{\rm for}\hspace{0.20cm}m_n^h \leq m_n \, ,
\nonumber \\
q_{h\,n}^- & \in & [0,\pi (m_n - m_n^h)] \, ,
\nonumber \\
q_{h\,n}^+ & = & q_{h\,n}^- + 2\pi m_n^h \, , \hspace{0.20cm}{\rm where}\hspace{0.20cm} n = 1,...,\infty \, ,
\label{M1compactH}
\end{eqnarray}
and
\begin{eqnarray}
M_n^0 (q_j) & = & \Theta (q_{p\,n}^+-q_j)\Theta (q_j-q_{p\,n}^-) \, , \hspace{0.20cm}{\rm for}\hspace{0.20cm}m_n^h \geq m_n \, ,
\nonumber \\
q_{p\,n}^- & \in & [0,-\pi (m_n - m_n^h)] \, ,
\nonumber \\
q_{p\,n}^+ & = & q_{p\,n}^- + 2\pi m_n \, , \hspace{0.20cm}{\rm where}\hspace{0.20cm} n = 1,...,\infty \, ,
\label{M1compactP}
\end{eqnarray}
respectively. The distribution $\Theta (x)$ in these equations is given by $\Theta (x)=1$ for $x\geq 0$ and $\Theta (x)=0$ for $x<0$.
For each LWS the $n$-band hole numbers $M_n^h$ appearing here are given in Eq. (\ref{Mh}). In each
LWS fixed-$S$ reduced subspace the set $\{M_n\}$ of numbers $M_n$ obey the two exact sum rules, Eqs. (\ref{Nsingletpairs})
and (\ref{Nps}), respectively. As given in Eqs. (\ref{M1compactH}) and (\ref{M1compactP}),
these hole-like and pseudoparticle-like general distributions refer to occupied $n$-bands 
of the LWSs under consideration for which $m_n^h\leq m_n$ and $m_n^h\geq m_n$, respectively.

LWSs for which $M_n^0 (q_j)=0$ for $n>1$,
$q_{p\,1}^+=-q_{p\,1}^-= \pi m_1$, and $m_1^h = m_S = m$ are 
ground states of the spin-$1/2$ $XXX$ chain Hamiltonian at finite magnetic field $h>0$, Eq. (\ref{H-h}). 
In that case their $n=1$ band momentum distribution, $M_1^0 (q_j) =M_1^{GS} (q_j)$,
refers to a compact pseudoparticle symmetrical distribution. Specifically, in the TL it reads $M_1^{GS} (q_j)=1$ for $q_j\in [-\pi m_1,\pi m_1]$ 
and thus is unoccupied, $M_1^{GS} (q_j)=0$, for $\vert q_j\vert \in [\pi m_1,\pi m_1^b]$.
As for all LWSs with symmetrical $n$-band distributions, 
ground states carry in the TL zero spin current. 

The large class of LWSs with compact $n$-band 
distributions, Eqs. (\ref{M1compactH}) and (\ref{M1compactP}), 
carry currents whose absolute value ranges from zero, for symmetrical compact distributions, 
to the corresponding LWS fixed-$S$ reduced subspace largest such values, Eq. (\ref{Jzmax}), for well-defined 
asymmetrical compact distributions in the $n$-bands with finite occupancy.

It is useful to consider the subspaces spanned by a given $S>0$ reference LWS 
with $n$-band compact distribution of the general form, Eqs. (\ref{M1compactH}) and (\ref{M1compactP}), and the set
of LWSs generated from it by processes involving pseudoparticle number overall 
deviations $\delta M_{\rm ps}$ for all $n=1,...,\infty$ bands 
such that $\delta m_{\rm ps}=\delta M_{\rm ps}/L\rightarrow 0$ as $L\rightarrow\infty$. 
Here $\delta M_{\rm ps}=\sum_{n=1}^{\infty}\delta M_n$ and thus
$\delta m_{\rm ps}=\sum_{n=1}^{\infty}\delta m_n$. 

A functional expression for the energy deviation $\delta E = E_{f} - E_0$, where $E_0$ stands for the energy of
the reference LWS and $E_{f}$ that of the LWSs generated from it, 
is derived from the use of the TBA energy spectrum, 
\begin{equation}
E = - \sum_{n=1}^{\infty}\sum_{j=1}^{M_n^b}M_{n} (q_{j})\,{J\over n}
\left(1+ \cos k^n (q_j)\right) - 2\mu_B\,h\,S^z \, .
\label{E-k}
\end{equation}
This is achieved upon expanding the excited states $n=1,...,\infty$ band momentum distributions 
$M_n (q_j) = M_n^0 (q_j) + \delta M_n (q_j)$ around $M_n^0 (q_j)$. Here the deviations
$\delta M_n (q_j) = M_n (q_j) - M_n^0 (q_j)$ and $\sum_{n=1}^{\infty}\sum_{j=1}^{M_n^b}\delta M_n (q_j)
= \delta M_{\rm ps}$ are as given above such that
$\delta m_{\rm ps}=\delta M_{\rm ps}/L\rightarrow 0$ as $L\rightarrow\infty$. Up to ${\cal{O}} (1/L)$ order 
one then finds,
\begin{equation}
\delta E = \sum_{n=1}^{\infty}\sum_{j=1}^{M_n^b}\varepsilon_n (q_j)\delta M_n (q_j) 
+ {1\over L}\sum_{n,n'=1}^{\infty}
\sum_{j,j'=1}^{M_n^b}{1\over 2}\,f_{n\,n'} (q_j,q_{j'})\delta M_n (q_j)\delta M_{n'} (q_{j'}) \, .
\label{DEnHm}
\end{equation}
Here the $n$-pseudoparticle dispersion $\varepsilon_n (q_j)$ reads,
\begin{equation}
\varepsilon_n (q_j) = 2n\mu_B\,h  -{J\over n}\Bigl(1 + \cos k^n_0 (q_j) 
- \sum_{n'=1}^{\infty}\int_{-\pi}^{\pi} dk \,{\bar{M}}_{n'}^0 (k) \,\sin k\,{\bar{\Phi}}_{n'\,n} (k,k^n_0 (q_j))\Bigr) \, .
\label{varepsilon-nHm}
\end{equation}
Here $k^n_0 (q_j)$ denotes the reference LWS momentum rapidities $k^n (q_j)$. Those are the solution of the TBA equations, Eq. (\ref{gen-Lambda}), 
for the compact $n$-band distributions, $M_{n} (q_{j})=M_{n}^{0} (q_{j})$, Eqs. (\ref{M1compactH}) and (\ref{M1compactP}),
of the reference state under consideration. The rapidity-variable distribution ${\bar{M}}_{n'}^0 (k)$ in Eq. (\ref{varepsilon-nHm})
is defined by  the relation ${\bar{M}}_{n'}^0 (k^{n'}_0 (q_j))= M_{n'}^0 (q_j)$.

The $f$ functions in Eq. (\ref{DEnHm}) are given by,
\begin{eqnarray}
f_{n\,n'} (q_j,q_{j'}) & = & v_n (q_{j})\,2\pi \,\Phi_{n\,n'} (q_{j},q_{j'}) 
+ v_{n'} (q_{j'})\,2\pi \,\Phi_{n'\,n} (q_{j'},q_{j}) 
\nonumber \\
& + & {1\over 2\pi}\sum_{n''=1}^{\infty}\sum_{\iota=\pm}\vert v_{n''} (q_{a\,n''}^{\iota})\vert\,2\pi\,\Phi_{n''\,n} (q_{a\,n''}^{\iota},q_{j})
\,2\pi\,\Phi_{n''\,n'} (q_{a\,n''}^{\iota},q_{j'}) \, , \hspace{0.25cm}{\rm for}\hspace{0.20cm}a = p,h \, .
\label{ffnHm}
\end{eqnarray}
The quantities $q_{a\,n''}^{\pm }$ where $a = p,h$ are here the compact distributions limiting momentum values
in Eqs. (\ref{M1compactH}) and (\ref{M1compactP}). Within the TL the $n$-band group velocity in
that expression reads,
\begin{equation}
v_n (q_j) = v_n (q)\vert_{q=q_j} \, ; \hspace{0.50cm} 
v_n (q) = {\partial\varepsilon_n (q)\over\partial q} \, .
\label{vn}
\end{equation}

Moreover, the rapidity dressed phase shifts ${\bar{\Phi}}_{n\,n'} (k,k')$ and related
momentum dressed phase shifts $\Phi_{n\,n'} (q_j,q_{j'})$ in units of $2\pi$ appearing both in the 
$\varepsilon_n (q_j)$ and $f_{n\,n'} (q_j,q_{j'})$ expressions are defined by the 
following integral equations and relations,
\begin{eqnarray}
{\bar{\Phi}}_{n\,n'} (k,k') & = & {1\over 2\pi}\Theta_{n\,n'}\left(n \tan (k/2) -n' \tan (k'/2)\right)
\nonumber \\
& - & \sum_{n''=1}^{\infty}{n''\over 4\pi}\int_{-\pi}^{\pi} dk'' \,{\bar{M}}_{n''}^0 (k'') 
{\Theta^{[1]}_{n\,n''} \left(n \tan (k/2) - n'' \tan (k''/2)\right)\over\cos^2 (k''/2)}\,
{\bar{\Phi}}_{n''\,n'} (k'',k') \, ,
\nonumber \\
\Phi_{n\,n'} (q_j,q_{j'}) & = & {\bar{\Phi}}_{n\,n'} (k^n_0 (q_j),k^{n'}_0 (q_{j'})) \, ,
\label{PsEqHm}
\end{eqnarray}
respectively. Here $\Theta_{n\,n'}(x)= - \Theta_{n\,n'}(-x)$ is the function given in 
Eq. (\ref{Theta}) and $\Theta^{[1]}_{n\,n'}(x)=\Theta^{[1]}_{n\,n'}(-x)$ 
is its derivative,
\begin{eqnarray}
\Theta^{[1]}_{n\,n'}(x) & = & \delta_{n,n'}\Bigl\{{1\over n\,(1+({x\over 2n})^2)}
+ \sum_{l=1}^{n -1}{2\over l(1+({x\over 2l})^2)}\Bigr\} 
\nonumber \\
& + & (1-\delta_{n,n'})\Bigl\{{2\over |n-n'|(1+({x\over
|n-n'|})^2)} 
\nonumber \\
& + & \sum_{l=1}^{{n+n'-|n-n'|-2\over 2}}{4\over
(|n-n'|+2l)(1+({x\over |n-n'|+2l})^2)} 
\nonumber \\
& + & {2\over (n+n')(1+({x\over n+n'})^2)}\Bigr\} \, .
\label{The1}
\end{eqnarray}

The two methods used in Ref. \cite{Carmelo-92} for the 1D Hubbard model
and in Ref. \cite{Bares-92} for the related $t-J$ model to calculate the elementary spin current 
$j_n (q_j)$, Eq. (\ref{jn-fn}), for reference LWSs with ground-state compact distributions
by means of conservation laws and under twisting boundary conditions, respectively, 
apply as well to the present more general compact distributions,
Eqs. (\ref{M1compactH}) and (\ref{M1compactP}). For
the spin-$1/2$ $XXX$ Heisenberg chain both such methods lead 
to exactly the same expression,
\begin{equation}
j_n (q_j) = - 2n\,v_{n} (q_j) -  \sum_{n'=1}^{\infty} {\iota_a\,n'\over \pi} \sum_{\iota =\pm} (\iota)f_{n\,n'} (q_j,q_{a\,n'}^{\iota}) \, ,
\label{jnvf}
\end{equation}
where $a = p,h$ and,
\begin{equation}
\iota_p = 1 \, ; \hspace{0.5cm} \iota_h = -1   \, .
\label{iotaHP}
\end{equation}

There are two limits in which the classes of LWSs 
considered here correspond to {\it all} existing such states: (i)
$(1-m_1^h) \ll 1$ when $(m_1^h-m_S) \ll 1$ and (ii) $m_1^h \ll 1$, respectively. In these two limiting cases
the use of elementary current, Eq. (\ref{jnvf}), gives for the $n=1$ and $n>1$ bands,
\begin{eqnarray}
j_1 (q_j) & = & - v_{1} (q_j) \, , \hspace{0.20cm}{\rm for}\hspace{0.20cm}m_1^h \ll 1 
\nonumber \\
& = & - 2\,v_{1} (q_j) \, , \hspace{0.20cm}{\rm for}\hspace{0.20cm}(1-m_1^h) \ll 1 \, ,
\nonumber \\
j_n (q_j) & = & - 2(n-1)\,v_{n} (q_j)
\, , \hspace{0.20cm}{\rm for}\hspace{0.20cm}m_1^h \ll 1  
\nonumber \\
& = &  - 2n\,v_{n} (q_j) \, , \hspace{0.20cm}{\rm for}\hspace{0.20cm}(1-m_1^h) \ll 1 \, ,
\label{jS1nrelav}
\end{eqnarray}
respectively. The $n$-band group velocities,
Eq. (\ref{vn}), in these expressions have the following exact behaviors,
\begin{eqnarray}
v_{1} (q_j) & = & J{\pi\over 2}\sin (q_j) \, , \hspace{0.20cm}{\rm for}\hspace{0.20cm}m_1^h \ll 1 \, ,
\nonumber \\
v_{n} (q_j) & = & J{2 (\pi m_1^h)^2\over 3n}\sin \left({q_j\over m_n^b}\right) \, , \hspace{0.20cm}
{\rm for}\hspace{0.20cm} n>1  \, , \hspace{0.15cm} m_S \ll 1 \, ,
\nonumber \\
v_{n} (q_j) & = & J\sin (q_j) \, , \hspace{0.20cm}{\rm for}\hspace{0.20cm}(1-m_1^h) \ll 1 \, .
\label{vnlimitsR}
\end{eqnarray}
By combining the relations, Eq. (\ref{jS1nrelav}), with the limiting group-velocity expressions
provided in Eq. (\ref{vnlimitsR}) one arrives to the elementary current $j_n (q_j)$ 
expressions, Eqs. (\ref{j1qj}) and (\ref{jS1n}), which is one of the goals of this Appendix.

\subsection{Elementary currents for LWSs described only by groups of real rapidities}
\label{ncurrentsR}

The goal of this Appendix section is to justify the validity of the inequality, Eq. (\ref{ineq2}).
It refers to the model in the LWS fixed-$S$ real-rapidity reduced subspaces
considered in Secs. \ref{CurrentsUB} and \ref{stiffnessHTh0}.

For the class of LWSs described only by groups of real rapidities and generated from reference states with
compact particle or hole $n=1$ band distributions, Eqs. (\ref{M1compactH}) and (\ref{M1compactP}), the general elementary current
expression, Eq. (\ref{jnvf}), simplifies for $n=1$ to,
\begin{equation}
j_1^S (q_j) = - 2v_{1} (q_j) -  {\iota_a\over \pi} \sum_{\iota =\pm} (\iota)f_{1\,1} (q_j,q_{a\,1}^{\iota}) \, .
\label{j1nvfRR}
\end{equation}
Here $a=p,h$ and the compact distribution limiting $n=1$ band momentum values belong to the
following intervals $q_{p\,1}^- \in [0,\pi (3m_S-1)/2]$ and $q_{p\,1}^+ = q_{p\,1}^- + \pi (1-m_S)\in [\pi (1-m_S),\pi (1+m_S)/2]$ for $a=p$ and
$q_{h\,1}^- \in [0,\pi (1-3m_S)/2]$ and $q_{h\,1}^+ = q_{h\,1}^- + 2\pi m_S\in [2\pi m_S,\pi (1+m_S)/2]$ for $a=h$.
This applies to the elementary current $j_1^S (q_j)$ in the current operator expectation value
$\langle\hat{J}^z (l_S,S)\rangle=\sum_{j =1}^{M_S^b} M_S^p (q_j)\,j_1^S (q_j)$, Eq. (\ref{J-partS}).

An interesting property refers to LWSs belonging to the fixed-$S$ real-rapidity reduced subspaces with symmetrical 
compact $n=1$ band distributions. For the present real-rapidity reduced subspaces, such LWSs
are actually $S>0$ ground states. They are a subclass of the LWSs with compact $n=1$ band distributions,
Eqs. (\ref{M1compactH}) and (\ref{M1compactP}) for $n=1$. These ground states
carry in the TL zero spin current. This follows from their elementary current being an odd function,
$j_1^S (q_j) = - j_1^S (-q_j)$. However, their elementary current absolute values reach the largest values.
The latter property renders the ground-state  elementary currents important for our analysis. 

One finds the following expressions for the corresponding ground-state $n=1$ group velocity for the 
whole range $m_S = m \in [0,1]$,
\begin{equation}
v_{1} (q_j) \approx \gamma_1^S J\,{\sin\left({\pi\over 2}m_S\right)\over m_S}\,\sin (q_j) \, , 
\label{vnlimitsR1}
\end{equation}
where 
\begin{equation}
\gamma_1^S = \sqrt{1-m_S(1-m_S)} \, .
\label{gamma1}
\end{equation}
The $v_{1} (q_j)$ expression given here is exact both for $m_S \ll 1$ and $(1-m_S) \ll 1$
and an excellent quantitative approximation for $m_S\approx 1/2$.

For such $S>0$ ground states the corresponding $n=1$ band elementary current reads,
\begin{equation}
j_{1}^S (q_j) = - 2(\xi^1)^2\,v_{1} (q_j) 
\approx - 2\gamma_1 J (\xi^1)^2\,{\sin\left({\pi\over 2}m_S\right)\over m_S}\,\sin (q_j) \, .
\label{currentsSymmR}
\end{equation}
In the TL the relation $j_{1}^S (q_j) = 2(\xi^1)^2\,v_{1} (q_j)$ is exact. The $j_{1}^S (q_j)$
expression given here is exact both for $m_S \ll 1$ and $(1-m_S) \ll 1$. 
For intermediate $m_S\approx 1/2$ densities it has an absolute value
$\vert j_{1}^S (q_j)\vert$ slightly larger than the corresponding exact value. 
Hence it is a very good approximation for the whole $m_S \in [0,1]$ range. 

The parameter $\xi^1$ in Eq. (\ref{currentsSymmR}) can be expressed
in terms of phase shifts $\Phi_{1\,1} (q_j,q_{j'})$ (in units of $2\pi$) defined by 
Eq. (\ref{PsEqHm}) for $n=n'=1$ as follows,
\begin{equation}
\xi^1 = 1 +  \Phi_{1\,1} (\pi m_1,\pi m_1^-) - \Phi_{1\,1} (\pi m_1,-\pi m_1) \, .
\label{xi0}
\end{equation}
Here $m_1 = (1-m_S)/2$ and $\pi m_1^-= \pi m_1 - 2\pi/L$. The parameter $\xi^1$ smoothly changes from 
$\xi^1 = 1/\sqrt{2}$ for $m_S\rightarrow 0$ to $\xi^1 = 1$ as $m_S\rightarrow 1$. 

\begin{figure}
\begin{center}
\centerline{\includegraphics[width=7.00cm]{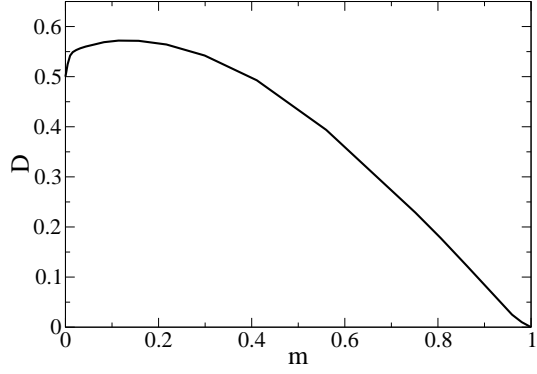}}
\caption{The zero-temperature spin stiffness $D=D(0)$, Eq. (\ref{chiD}), 
plotted in units of $J=1$ as a function of the spin density $m=m_S\in [0,1]$.}
\label{figure2}
\end{center}
\end{figure}

Since the present symmetrical compact LWSs are ground states, one finds that 
the dressed phase-shift parameter $\xi^1$, Eq. (\ref{xi0}), is 
directly related to the model zero-temperature spin stiffness, $D=D(0)$. Indeed, 
the elementary current absolute value $\vert j_1^S (\pi m_1)\vert = 2(\xi^1)^2\, v_1 (\pi m_1)$ at
$q_j = \pi m_1 = \pi (1-m_S)/2 = \pi (1-m)/2$
fully controls such a zero-temperature stiffness for $m=m_S\in [0,1]$ as follows \cite{CPJD-15},
\begin{equation}
\pi D(0) = 2(\xi^1)^2\, v_1^S (\pi m_1) = \vert j_1^S (\pi m_1)\vert \, .
\label{chiD}
\end{equation}
The dependence on $m=m_S$ of the zero-temperature spin stiffness, Eq. (\ref{chiD}), has been 
investigated in previous studies \cite{CPJD-15}. It is plotted in Fig. \ref{figure2}.

The elementary currents $j_1^S (q_j)$ of all fixed-$S$ LWSs 
described only by groups of real rapidities can be written as,
\begin{equation}
j_1^S (q_j) = j_1^S\,s_1^S (q_j) \, ; \hspace{0.5cm} \vert s_1^S (q_j)\vert \leq 1 \, ,
\label{j1Sgen}
\end{equation}
where $j_1^S>0$ is the largest elementary current absolute value.

As mentioned above, although the class of LWSs with 
symmetrical compact distributions, Eqs. (\ref{M1compactH}) and (\ref{M1compactP}), carry zero current,
their elementary currents absolute values reach the largest values of each $S$-fixed subspace.
The largest absolute value $j_1^S = \vert j_1^S (q_{w})\vert$
of the symmetrical compact distribution ground-state 
elementary current, Eq. (\ref{currentsSymmR}), is reached at $q_j=q_{w}\approx \pm \pi/2$ and
reads $\pi D(0) v_1 (q_{w})/v_1  (\pi m_1)\approx \pi D(0)/\cos\left(\pi  m_S/2\right)$. It  is 
a continuous increasing function of $m_S$ that smoothly varies from its minimum value $J\pi/2$ 
for $m_S\rightarrow 0$ to its maximum value $2J$ as $m_S\rightarrow 1$. Moreover, for {\it all} fixed-$S$ 
LWSs described only by groups of real rapidities the following two universal limiting behaviors hold,
\begin{equation}
\lim_{m_S\rightarrow 0} j_1^S = J{\pi\over 2} \, ; \hspace{0.5cm}
\lim_{m_S\rightarrow 1} j_1^S = 2J \, .
\label{limitsjmax}
\end{equation}

Manipulations for intermediate $m_S \in [0,1]$ values of the BA equations, Eq. (\ref{gen-Lambda-BA}) and
Eq. (\ref{gen-Lambda}) for $n=1$, and spin/$n=1$ band elementary current expressions, 
Eq. (\ref{j-k}), Eq. (\ref{jn-fn}) for $n=1$, and Eq. (\ref{j1Sgen}), confirms that, as for the ground 
state, the largest elementary current absolute value $j_1^S$ of 
all the LWSs described only by groups of real rapidities is for $m_S<1$ smaller than $2J$. Hence,
\begin{eqnarray}
{\pi D(0)\,v_1 (q_{w})\over v_1  (\pi m_1)} & < & 2J \, , \hspace{0.20cm}
{\rm for}\hspace{0.20cm}m_S < 1 \, ,
\nonumber \\
j_1^S  & < & 2J \, , \hspace{0.20cm}{\rm for}\hspace{0.20cm}m_S < 1 \, .
\label{D0max}
\end{eqnarray}
The first inequality refers to the largest elementary current absolute value 
$j_1^S=\pi D(0) v_1 (q_{w})/v_1  (\pi m_1)\approx \pi D(0)/\cos\left({\pi\over 2} m_S\right)$
reached for $S>0$ ground states. It has been expressed in terms of the zero-temperature 
spin stiffness for $m=m_S$. The second inequality in Eq. (\ref{D0max}) applies to 
the largest elementary current absolute value $j_1^S$ of all LWSs with fixed spin $S$ that are 
described only by groups of real rapidities. 

The limiting behaviors, Eq. (\ref{limitsjmax}), and inequalities, Eq. (\ref{D0max}), justify the largest elementary 
current absolute value $j_1^S=2J$ of the elementary current $j (q_j) = -2J\,\sin q_j$, Eq. (\ref{jn-fn0}), used in our 
$T\rightarrow\infty$ spin stiffness upper bound scheme of Sec. \ref{stiffnessHTh0}. Next, we briefly describe the main
mechanism that justifies the use of the function $s_1^S (q_j)=-\sin q_j$. First we discuss the suitable use
of a odd function, $s_1^S (q_j)=-s_1^S (-q_j)$, for that elementary current. We then justify the specific
choice, $s_1^S (q_j)=-\sin q_j$.  

On the one hand, that we use a odd function for $s_1^S (q_j)$ is all right for LWSs with symmetrical compact
and symmetrical non-compact distributions such that $M_S^p (q_{j})=M_S^p (-q_{j})$. 
On the other hand, analysis of the BA equation reveals that the exact function 
$s_1^S (q_j)$ in Eq. (\ref{j1Sgen}) such that $\vert s_1^S (q_j)\vert \leq 1$
is not a odd function of $q_j$ for general LWSs with asymmetrical  compact and asymmetrical non-compact distributions
such that $M_S^p (q_{j})\neq M_S^p (-q_{j})$. Nonetheless, 
the use of a odd function $s_1^S (q_j)$ for these states enhances in general their
current absolute values, $\vert\sum_{j =1}^{M_S^b} M_S^p (q_j)\,j_1^S (q_j)\vert$.

Our following analysis applies to general LWSs described only by groups of real rapidities. Those do not
necessarily have compact $M_S^p (q_{j})$ occupancies. Hence rather than the elementary current
$j_1^S (q_j)$ given in Eq. (\ref{j1nvfRR}), which is specific to such occupancies, here we use
the more general elementary current $j_1^S (q_j) = - {2J\sin k^1 (q_j)\over 2\pi\sigma_1 (k^1 (q_j))}$.
It is that given in Eq. (\ref{jn-fn}) for $n=1$ and LWSs described only by groups of real rapidities.

For all such LWSs the BA equation is of the same form, Eq. (\ref{gen-Lambda-BA})
and Eq. (\ref{gen-Lambda}) for $n=1$, for large finite $L$ and the TBA, respectively. It can
be written as,
\begin{equation}
q_j = k^1 (q_j) - {2\over L}\sum_{j'=1}^{M_S^b} M_S^p (q_{j'})\,\arctan \left({\tan (k^1 (q_j)/2) - \tan (k^1 (q_{j'})/2)\over 2}\right) , 
\label{kjsolu}
\end{equation} 
where $j = 1,...,M_S^b$. If the momentum distribution is an even function, $M_S^p (q_{j'})=M_S^p (-q_{j'})$,
one finds that $k^1 (0)=0$ at $q_j=0$. The elementary current, $j_1^S (q_j) = - {2J\sin k^1 (q_j)\over 2\pi\sigma_1 (k^1 (q_j))}$, 
is then a odd function. This follows from the distribution $2\pi\sigma_1 (k)$ turning out to be an even function in that case. The
latter distribution can be written as $2\pi\sigma_1^b (k)\,{\bar{M}}_S^p (k)$
and equivalently as $2\pi\sigma^b (k)\,{\bar{M}}_S^p (k)$. Here $2\pi\sigma_1^b (k)$
is the distribution, Eq. (\ref{sigmsrhoderiv}) for $n=1$, and $2\pi\sigma^b (k_j)$ is the solution of 
Eq. (\ref{sigma-IE-n1}). For the present case of real rapidities they are the same distributions. Moreover,
${\bar{M}}_S^p (k_j)=M_S^p (q_j)$.

In the general case of LWSs for which the momentum distribution $M_S^p (q_{j})$ is not an even function,
$M_S^p (q_{j})\neq M_S^p (-q_{j})$, the corresponding elementary current $j_1^S (q_j)$ is not a odd function. 
Consistently, the $n=1$ band momentum $q_j=0$ then corresponds to a finite momentum rapidity $k^1 (0)$ given by,
\begin{equation}
k^1 (0) = {2\over L}\sum_{j=1}^{M_S^b} M_S^p (q_{j})\,
\arctan \left({\tan (k^1 (q_{j})/2)+\tan (k^1 (0)/2)\over 2}\right) \, ,
\label{k0}
\end{equation} 
such that $k^1 (0) <  \pi (1-m_S)/2$.

This implies that there is a positive or negative $q_j$ interval, 
\begin{eqnarray}
& & q_j \in [0,q^0] \rightarrow k^1_j \in [-k^1 (0),k^1 (0)] \hspace{0.5cm}{\rm for}\hspace{0.2cm}q^0 >0
\nonumber \\
& & q_j \in [q^0,0] \rightarrow k^1_j \in [-k^1 (0),k^1 (0)] \hspace{0.5cm}{\rm for}\hspace{0.2cm}q^0 <0 \, ,
\label{corre}
\end{eqnarray} 
where,
\begin{equation}
q^0 = {2\over L}\sum_{\iota =\pm1}\sum_{j=1}^{M_S^b} M_S^p (q_{j})\,\arctan \left({\tan (k^1 (q_{j})/2)+(\iota)\tan (k^1 (0)/2)\over 2}\right) \, ,
\label{q0}
\end{equation} 
in which the elementary current, $j_1^S (q_j) = - {2J\sin k^1 (q_j)\over 2\pi\sigma_1 (k^1 (q_j))}$, has
opposite signs for the two subintervals $k^1_j \in [-k^1 (0),0]$ and $k^1_j \in [0,k^1 (0)]$, respectively.
This refers to the corresponding momentum rapidity interval $k^1_j \in [-k^1 (0),k^1 (0)]$. Indeed
the distribution $2\pi\sigma_1 (k^1 (q_j))=2\pi\sigma_1 (k^1_j)$ is for all LWSs such that 
$2\pi\sigma_1 (k^1_j)\geq 0$. And this applies to its whole range $k^1_j\in [-\pi,\pi]$ and thus corresponding $q_j$ 
range $q_j \in \left[-\pi (1-m_S)/2,\pi (1-m_S)/2\right]$.

In the $q_j$ interval $q_j \in [0,q^0]$ for $q^0>0$ and $q_j \in [q^0,0]$ for $q^0<0$ the 
band momentum $q_j$ has the same sign. However, the elementary current $j_1^S (q_j)$ has opposite signs 
in two momentum $q_j$ subintervals of these intervals. For example, for $q^0>0$ such subintervals read 
$q_j \in [0,q (0) ]$ and $q_j \in [q (0),q^0]$, respectively. Here,
\begin{equation}
q (0) = {2\over L}\sum_{j=1}^{M_S^b} M_S^p (q_{j})\arctan \left({\tan (k^1 (q_{j})/2)\over 2}\right) \, ,
\label{q*}
\end{equation} 
is the $q_j$ value at which the momentum rapidity vanishes, $k^1 (q (0)) =0$. 

The function $s_1^S (q_j)$ in Eq. (\ref{j1Sgen}) such that $\vert s_1^S (q_j)\vert \leq 1$ has the same signs as $k^1_j$. 
It follows that the current contributions from occupancies in such $q^0>0$ subintervals, $q_j \in [0,q (0)]$ and $q_j \in [q (0),q^0]$, tend to cancel. 
This would not be so if $s_1^S (q_j)$ was a odd function. Moreover, the cancelling momentum rapidity interval $k^1_j \in [-k^1 (0),k^1 (0)]$
corresponds to $q_j$ alternative positive $q_j \in [0,q^0]$ and negative $q_j \in [q^0,0]$ intervals if the asymmetric 
distribution $M_S^p (q_{j})$ has integrated larger values for $q_j>0$ and $q_j<0$, respectively. Hence the use of a suitably chosen odd
function $s_1^S (q_j)$ enhances indeed the current absolute values $\vert\sum_{j =1}^{M_S^b} M_S^p (q_j)\,j_1^S (q_j)\vert$
of most LWSs.

Finally, we justify the choice of the specific odd function, $s_1^S (q_j)=-\sin q_j$. As follows from Eq. (\ref{j1qj}) 
for $m_1^h = m_S$, one finds for {\it all} LWSs described only by groups of real rapidities 
that in the limits $m_S\ll 1$ and $(1-m_S)\ll 1$ their elementary currents $j_1^S (q_j)$ are exactly given by,
\begin{eqnarray}
j_1^S (q_j) & = & - J{\pi\over 2}\sin (q_j) \, , \hspace{0.20cm}{\rm for}\hspace{0.20cm}m_S \ll 1 
\nonumber \\
& = & - 2J\,\sin (q_j) \, , \hspace{0.20cm}{\rm for}\hspace{0.20cm}(1-m_S) \ll 1 \, ,
\label{j1qjmS}
\end{eqnarray}
respectively. The simplest odd function $s_1^S (q_j)=-s_1^S (-q_j)$ that
in these two limits reaches the exact behavior of the elementary currents carried by such LWSs 
is indeed $s_1^S (q_j)=-\sin q_j$. Additionally, we have confirmed
that this choice enhances the current absolute values $\vert\sum_{j =1}^{M_S^b} M_S^p (q_j)\,j_1^S (q_j)\vert$
of most LWSs. Importantly, it enhances in {\it all} LWS fixed-$S$ subspaces under
consideration the quantity $\sum_{l_{\rm r_S}}\vert\langle\hat{J}^z (l_{\rm r_S},S)\rangle\vert^2/(2S)^2$ on the
right-hand side of Eq. (\ref{D-T-u1}).

Combining all the above arguments and properties justifies why the replacement of the exact elementary functions $j_1^S (q_j)$
by $j (q_j) = -2J\,\sin q_j$, Eq. (\ref{jn-fn0}), in the current absolute values 
$\vert\sum_{j =1}^{M_S^b} M_S^p (q_j)\,j_1^S (q_j)\vert$ of all LWSs described only by groups of real rapidities
leads to the inequality, Eq. (\ref{ineq2}).

\section{Derivation of the second stiffness upper bound}
\label{currentUB2}

Here the sum $\sum_{l_{\rm r_1}}$ in Eq. (\ref{D-T-u1}) is performed by the use of the upper-bound elementary spin current
$j (q_j) = -2J\,\sin q_j$, Eq. (\ref{jn-fn0}). To reach this goal we first consider the $M_S^p$-dependent sums, for fixed $M_S^b$.
Those give a upper bound on corresponding sums over $l_S$ in Eq. (\ref{D-T-u1}), namely, 
\begin{equation}
\frac{1}{4J^2}\sum_{l_S}\vert\langle\hat{J}^z (l_S,S)\rangle\vert^2 \le I (M_S^p) \, .
\label{est1}
\end{equation}
Here,
\begin{equation}
I(M_s^p) = \sum_{b_1,b_2\ldots b_{M^b_S}\in\{ 0,1\}} \delta_{M_S^p,\sum_l b_l} \left|\sum_{k=1}^{M^b_S} b_k \sin q_k\right|^2 \, ,
\end{equation}
and $b_j\equiv M(q_j)$ are binary occupation numbers, which we sum over.

The $\delta$-constrain can be analytically treated by means of a counting field parameter $\lambda$. 
This is done by defining,
\begin{equation}
\tilde{I}(\lambda) = \sum_{M_S=0}^{M_S^b} e^{\lambda M_S^p} I(M_S^p) \, .
\end{equation}
We then find immediately that, 
\begin{eqnarray}
\tilde{I}(\lambda) &=& \sum_{k=1}^{M^b_S}\sum_{l=1}^{M^b_S}\sin q_k\sin q_l\, \sum_{b_1,b_2\ldots b_{M^b_S}} b_k b_l \prod_{j} e^{\lambda b_j} \nonumber \\
&=& \sum_k \sin^2 q_k\, e^\lambda (1+ e^\lambda)^{M_S^b-1} \nonumber \\ 
&&+ \sum_{k\neq l} \sin q_k \sin q_l\, e^{2\lambda} (1+e^\lambda)^{M_S^b-2} \nonumber \\
&=& \sum_k \sin^2 q_k\, (e^\lambda (1+ e^\lambda)^{M_S^b-1}- e^{2\lambda} (1+e^\lambda)^{M_S^b-2} ) \nonumber \\
&=&  \sum_k \sin^2 q_k\, e^\lambda (1+ e^\lambda)^{M_S^b-2} \nonumber \\
&=& \left(\sum_k \sin^2q_k\right)\sum_{M_S^p=1}^{M^b_S-1}{M^b_S -2 \choose M_S^p - 1} e^{M_S^p \lambda} \, .
\end{eqnarray}
Indeed, due to $\delta$-constraint one has that $e^{\lambda M_S^p} = \prod_{k=1}^{M^b_S} e^{\lambda b_k}$.

We have been using the property that $0 = (\sum b_k \sin q_k)^2 = \sum_k \sin^2 q_k + \sum_{k\neq l} \sin q_k \sin q_l$. 
From it we find $I(M_s^p)$,
\begin{eqnarray}
I(0) & = & I(M^b_S) = 0, \\
I(M_S^p) & = & \left(\sum_{k=1}^{M^b_S} \sin^2 q_k\right){M^b_S - 2 \choose M_S^p - 1} \nonumber \\
&\quad& {\rm for} \; 1 \le M_S^p \le M^b_S-1 \, .
\label{est1b}
\end{eqnarray}
Furthermore, we can explicitly calculate the sum over $\sin^2 q_k$. This gives,
\begin{eqnarray}
\sum_k \sin^2 q_k &=& \sum_{k=1}^{M^b_S} \sin^2\left(\frac{\pi}{L}(2k-M^b_S-1)\right) \nonumber \\
&=& \frac{1}{2}\left(M^b_S - \frac{\sin(2\pi M^b_S/L)}{\sin(2\pi/L)}\right) \nonumber \\
&=& \frac{L}{4}+\frac{S}{2}+\frac{1}{2}\frac{\sin(2\pi S/L)}{\sin(2\pi/L)} \, .
\label{est2}
\end{eqnarray}

From the use of the estimates in Eq. (\ref{est1}) with Eqs. (\ref{est1b}) and (\ref{est2}) in the expression for the stiffness,
Eq. (\ref{D-Tlarge}), we finally arrive at the simple bound given in Eq. (\ref{bound}), where a single sum over $S$ remains.

Next we confirm the behaviors reported in Eq. (\ref{value-D-m0-1}), which are reached  by the stiffness upper bound, Eq. (\ref{bound}),
in the $m\to 0$ and $m\to 1$ limits as $L\rightarrow\infty$. 
Concerning the $m\to 0$ limit,
within the TL one may replace ${L/2 + S - 2\choose L/2 - S - 1}$ on the right-hand side of Eq. (\ref{bound})
by a simpler expression, ${L/2 + S\choose L/2 - S}$. Hence the following identity can be used, 
\begin{eqnarray}
& & \sum_{S=0}^{L/2} {L/2+S\choose L/2-S} = f_{L+1} \, ,
\nonumber \\
& & \lim_{L\to\infty} {\left(\sum_{S=1}^{L/2-1}\varphi(S/L){L/2+S\choose L/2-S}\right)
\over\left(\sum_{S=0}^{L/2}{L/2+S\choose L/2-S}\right)}= \varphi(1/3) \, .
\label{sum}
\end{eqnarray}
Here $f_j$ is the $j-$th Fibonacci number, defined by $f_0 = f_1 = 1$, $f_{j+1} = f_j + f_{j-1}$, and $\varphi(x)$ is an arbitrary 
smooth function on $(0,1)$, possibly with poles at $0$ or $1$. In our case, 
\begin{equation}
\varphi(x) = \frac{1}{x^2}\left(\frac{1}{2} +  x + \frac{1}{2\pi}\sin(2\pi x)\right) \, .
\end{equation}
The replacement of ${L/2 + S - 2\choose L/2 - S - 1}$ by ${L/2 + S\choose L/2 - S}$ on the right-hand side of Eq. (\ref{bound})
amounts though by multiplying it by an additional factor,
\begin{equation}
\lim_{L\rightarrow\infty}{\sum_S {L/2 + S - 2\choose L/2 - S - 1}\over\sum_S { L/2 + S\choose L/2 - S }}
=  \lim_{L\rightarrow\infty} {f_{L-2}\over f_{L+1}} =   \sqrt{5} - 2 \, .
\label{extraF}
\end{equation}

From the combination of such procedures, we arrive at the following final compact upper bound valid for
$m=-2S^z/L \to 0$ in the TL,
\begin{equation}
D(T) \le  \frac{9}{2}(\sqrt{5} - 2)\left(\frac{5}{3} + \frac{\sqrt{3}}{2\pi}\right)\frac{J^2}{T} \left(\frac{S^z}{L}\right)^2 \, .
\label{ineqII}
\end{equation}
This is the expression given in Eq. (\ref{value-D-m0-1}) for $m\ll 1$. 

Note that the lower limit of the sum in Eq. (\ref{sum}) can in the TL be pulled up to 
the $S=\vert S_z\vert$ for any $\vert S_z\vert \le L/3$. This is so that the sum still starts before the maximum of the binomial 
symbol, which in the TL can be approximated with a gaussian This yields the same asymptotic inequality,
Eq. (\ref{ineq}).

Finally, we evaluate the behavior of the stiffness upper bound, Eq. (\ref{bound}), in the regime $m\to 1$, i.e., $-S^z = L/2 - \delta$, 
where $\delta \ll L$. This is a simple task fulfilled by using the leading order asymptotic in $\delta/L = 1-m$, which gives,
\begin{eqnarray}
D_{u2}(T)  &\simeq& 
\frac{J^2 (S^z)^2}{L T} \frac{\sum_{S=|S^z|}^{L/2-1} \frac{4}{L} {L/2+S - 2 \choose L/2 -S - 1}}{\sum_{S=|S^z|}^{L/2} {L/2+S\choose L/2-S}} 
\nonumber \\
&=& \frac{J^2 (S^z)^2}{L T} \frac{4}{L}  \frac{\sum_{k=1}^{\delta} {L-k-2 \choose k-1}}{\sum_{k=0}^{\delta} {L-k\choose k}}
\nonumber \\
&\simeq& \frac{J^2 (S^z)^2}{L T} \frac{4}{L} \frac{\sum_{k=0}^{\delta-1} L^k/k!}{\sum_{k=0}^{\delta} L^k/k!} 
\nonumber \\
&\simeq& \frac{4J^2}{T} \left(\frac{S^z}{L}\right)^2 \frac{\delta}{L} = \frac{J^2}{2T}(1-m) \, .
\end{eqnarray}
This is the behavior reported in Eq. (\ref{value-D-m0-1}) for $(1-m)\ll 1$.

\end{document}